\documentclass[10pt,twocolumn,twoside]{IEEEtran}
\usepackage{cite}
\usepackage{graphicx}
\usepackage{epstopdf}
\usepackage[cmex10]{amsmath}
\usepackage[caption=false,font=footnotesize]{subfig}
\usepackage{stfloats}
\usepackage{multirow}
\usepackage{url}
\usepackage{algorithm}
\usepackage{algorithmic}
\usepackage{amssymb}
\usepackage{amsmath}
\usepackage{balance}

\begin{document}
\title{Structure-Aware Bayesian Compressive Sensing\\ for Frequency-Hopping Spectrum Estimation\\ with Missing Observations}

\author{Shengheng~Liu,~\IEEEmembership{Member,~IEEE},
    Yimin~D.\ Zhang,~\IEEEmembership{Senior Member,~IEEE},
	Tao~Shan,~\IEEEmembership{Member,~IEEE},
	and~Ran~Tao,~\IEEEmembership{Senior Member,~IEEE}

\thanks{The work of S.-H.\ Liu, T.\ Shan, and R.\ Tao was supported in part by the National Natural Science Foundation of China under Grants Nos. 61671060, 61421001, 61331021, and Natural Science Foundation of Beijing Municipality under Grant No. 4172052. The work of Y.~D.\ Zhang was supported in part by the National Science Foundation under Grant No. AST-1547420. S.-H.\ Liu gratefully acknowledges the financial support from the China Scholarship Council for his stay at Temple University. Part of this work was presented at the 2016 IEEE Radar Conference, Philadelphia, PA \cite{liu16c1} and the 2016 SPIE Compressive Sensing Conference, Baltimore, MD \cite{liu16c2}.}
\thanks{S.-H.\ Liu was with the School of Information and Electronics, Beijing Institute of Technology, Beijing 100081 China, and also with the Department of Electrical and Computer Engineering, Temple University, Philadelphia, PA 19122 USA (e-mail: henry@bit.edu.cn).}
\thanks{Y.~D.\ Zhang is with the Department of Electrical and Computer Engineering, Temple University, Philadelphia, PA 19122 USA (e-mail: ydzhang@temple.edu).}
\thanks{T.\ Shan and R.\ Tao are with the School of Information and Electronics, Beijing Institute of Technology, Beijing 100081 China (e-mails: shantao@bit.edu.cn, rantao@bit.edu.cn).}
\thanks{Color versions of one or more of the figures in this paper are available online at http://ieeexplore.ieee.org.}
\thanks{Digital Object Identifier 10.1109/TSP.2018.XXXXXXX}
}

\markboth{IEEE TRANSACTIONS ON SIGNAL PROCESSING,~Vol.~XX, No.~X, XXX~XXXX}%
{Liu \MakeLowercase{\textit{et al.}}: STRUCTURE-AWARE BCS FOR FH SPECTRUM ESTIMATION WITH MISSING SAMPLES}

\maketitle

\begin{abstract}
In this paper, we address the problem of spectrum estimation of multiple frequency-hopping (FH) signals in the presence of random missing observations. The signals are analyzed within the bilinear time-frequency (TF) representation framework, where a TF kernel is designed by exploiting the inherent FH signal structures. The designed kernel permits effective suppression of cross-terms and artifacts due to missing observations while preserving the FH signal auto-terms. The kernelled results are represented in the instantaneous autocorrelation function domain, which are then processed using a re-designed structure-aware Bayesian compressive sensing algorithm to accurately estimate the FH signal TF spectrum. The proposed method achieves high-resolution FH signal spectrum estimation even when a large portion of data observations is missing. Simulation results verify the effectiveness of the proposed method and its superiority over existing techniques.
\end{abstract}


\section{Introduction}

\IEEEPARstart{F}{requency-hopping} (FH) signals are generated by varying the carrier frequencies according to a certain hopping pattern, which is typically pseudo-random. Due to their inherent capability of low probability of intercept, reduced interference to/from other users, resistance to jamming and multipath fading, and desirable ambiguity function property, FH signals have become a favorable choice in a wide range of communication and radar applications, particularly in the context of multiple-input multiple-output (MIMO) operations \cite{mari92, eusipco08 , chen08, hunt09, gogi12}. For a variety of tasks ranging from interception of non-cooperative emitters to exploitation of signals of opportunity for passive sensing, estimating and tracking the instantaneous spectrum of FH signals are an important yet challenging task when the hopping patterns of the constituent signals are unavailable. The problem becomes even more difficult when the hopping period is time-varying \cite{ange10}.

In this paper, we consider the spectrum estimation of multi-emitter FH signals with unknown and time-varying hopping periods in the context of Bayesian compressive sensing (BCS). In particular, we focus on the case where the received signal waveform is subject to missing observations. The specific FH signal structures are utilized to design time-frequency (TF) kernels and BCS structure priors to achieve reliable and high-resolution FH spectrum estimation.

The continuous-time noisy multi-emitter FH signal considered in this paper is expressed as \cite{simo95}
\begin{equation}\label{eq:fh_sig_conti}
s(t) = \sum\limits_{h = 1}^H {\sum\limits_{k = 1}^{{K_h}} {{A_{h,k}}\Pi _{T_h}(t - k{T_h}){{\rm{e}}^{{\rm{\jmath}}2\pi {f_{h,k}}(t - k{T_h})}}}  + v(t)},
\end{equation}
where $\jmath=\sqrt { - 1} $, and $\Pi _{T_h}(t)$ represents a normalized boxcar function, which is equal to one for $t \in \left( { - \frac{{{T_h}}}{2},\frac{{{T_h}}}{2}} \right]$ and $0$ otherwise. In addition, $T_h$ denotes the duration of each hop of the $h$-th individual FH emitter, and $H$ is the number of FH emitters. Moreover, $A_{h,k}$ and $f_{h,k}$  are the complex amplitude and carrier frequency of the $k$-th tone in the $h$-th system-wise dwell, respectively. The number of tones, $K_h$, may vary with $h$ because of emitter (de)activation or bandwidth mismatch \cite{ange10}. $v(t)$ represents the additive circularly-symmetric complex white Gaussian noise. Let $f_s$ and $\Delta t=1/{f_s}$ respectively denote the sampling rate and the sampling interval. Then, the sampled discrete-time FH signal can be derived from (\ref{eq:fh_sig_conti}) as
\begin{equation}\label{eq:fh_sig_discr}
s[n] = \sum\limits_{h = 1}^H {\sum\limits_{k = 1}^{{K_h}} {A_{h,k}}{{\rm{e}}^{{\rm{\jmath}}2\pi {f_{h,k}}n\Delta t}}  + v[n]}.
\end{equation}

In practice, the measured data may experience missing samples due to channel distortion/fading, line-of-sight obstruction, removal of samples contaminated by impulsive noise, and collecting/storage equipment failures \cite{amin15}. Denote $x[n]$ as signal $s[n]$ with missing data, and $N_m\subset \{1,2,\ldots,N\}$ as the set of missing time instants with cardinality $\left| {N_m} \right|=M$, where $N$ is the total length of signal $x[n]$ and  $s[n]$, and $M/N$ represents the missing-sample ratio. Then, $x[n]$ can be interpreted as $s[n]$ modulated by a sum of Dirac delta functions (impulses) \cite{Branka15}, i.e.,
\begin{equation}\label{eq:xn}
x[n] = s[n]\left( {1 - \sum\limits_{{n_m \in N_m}} {\delta \left[ {n - {n_m}} \right]} } \right).
\end{equation}
Random missing observations induce noise-like artifacts in the time-frequency distributions (TFD) \cite{zhan13}, which makes the problem even more intractable.

\subsection{Related Work}

Time-varying spectrum signatures of non-stationary signals, such as FH signals, can be revealed in the joint TF domain representations. As FH signals generally exhibit sparsity in the joint TF domain, compressive sensing (CS) and sparse reconstruction techniques \cite{dono06, emma06, Flan10} enable effective FH spectrum representation and parameter estimation. In \cite{ange10, ange13}, this problem is solved by formulating the problem as an underdetermined linear regression with a dual sparsity penalty, i.e., a penalty function that controls both the intrinsic sparsity and smoothness of the estimation. However, this approach requires appropriate tuning of the parameters, and obtaining the optimum solution still requires considerable effort. Another limitation of the approach proposed in \cite{ange10, ange13} is that they do not provide robust estimation performance due to the sensitivity of the differential operator used in fused least absolute shrinkage and selection operator (LASSO). To improve the parameter estimation performance, particularly in low signal-to-noise ratio (SNR) conditions, a BCS method was adopted in \cite{zhao15}, where a logistic stick breaking process is employed to encourage the temporal clustering over each hopping interval. BCS algorithm enables, through the proper design of priors, the incorporation of the contiguity property of typical TF signatures and thus enhances sparse optimization solutions. However, all the aforementioned approaches are based on linear TF analyses, and do not account for the effect of missing observations. Actually, linear approaches fail in the case of missing samples, as we will show in Section \ref{s:nea}.

As described in \cite{zhan13, jok14, stan14, amin15, Branka15}, the effect of artifacts induced by random missing samples can be substantially reduced by applying proper TF kernels, which involves developing FH spectrum estimation methods in the bilinear time-frequency representation (TFR) framework. Sparse reconstruction of TFRs using different CS methods can also be found therein. It is known that bilinear (quadratic) TFDs provide high-resolution time-varying spectrum representations. The Wigner-Ville distribution (WVD) is considered as a prototype of bilinear TFDs, which offers highest TF energy concentration for single-component linear frequency modulated signals. However, because of the bilinear nature, it causes cross-terms between different components that constitute false energy distributions. To resolve this problem, various reduced-interference distributions have been developed for cross-term reduction through the design of appropriate TF kernels in the general Cohen's class \cite{cohe95, boas15}. Such TF kernels can be signal-independent or signal-dependent. The latter performs parameter tuning via optimization, and thus generally provides better performance in trading off the cross-term suppression and the auto-term preservation. In particular, the adaptive optimal kernel (AOK), which is based on the optimization of radial Gaussian functions in the ambiguity function (AF) domain \cite{bara93, jone95}, is a commonly used signal-dependent kernel.

Recently, such approaches have been adopted to estimate FH spectrum from data with missing observations, and an orthogonal matching pursuit (OMP) algorithm based approach was developed to achieve both artifact mitigation and high-resolution FH signal spectrum estimation \cite{liu16c1}. The filtering capability of TF kernels offers bilinear TFR unique advantages over its linear counterpart \cite{ange10, ange13, zhao15} to effectively suppress the artifacts induced by missing observations. In the underlying problem that deals with FH spectrum estimation, however, separately reconstructing the TFR in each time instant as in \cite{liu16c1} does not utilize an important signal characteristic relevant to the contiguous structure of the FH signatures. In particular, the approach may likely generate isolated or sporadic entries in the reconstructed TFR in the presence of missing data and/or measurement noise. In \cite{wu14}, a novel continuous structure based BCS approach \cite{ji08} is proposed for the sparse reconstruction of nonstationary signals with missing observations. On this basis, a re-designed BCS-based scheme that exploits the contiguous structure of the FH signal is applied in \cite{liu16c2, liu18} to provide additional robustness in the FH spectrum estimation. Compared with the FH spectrum estimation via OMP \cite{liu16c1}, the BCS-based approach is proved capable to achieve an improved sparse solution \cite{ji08}. The BCS methods approach sparse solutions that are close to $\ell_0$-norm optimization and support structure-aware sparse problems through the use of adequate priors.

\begin{figure*}[!b]
\centering
\includegraphics[scale=0.45]{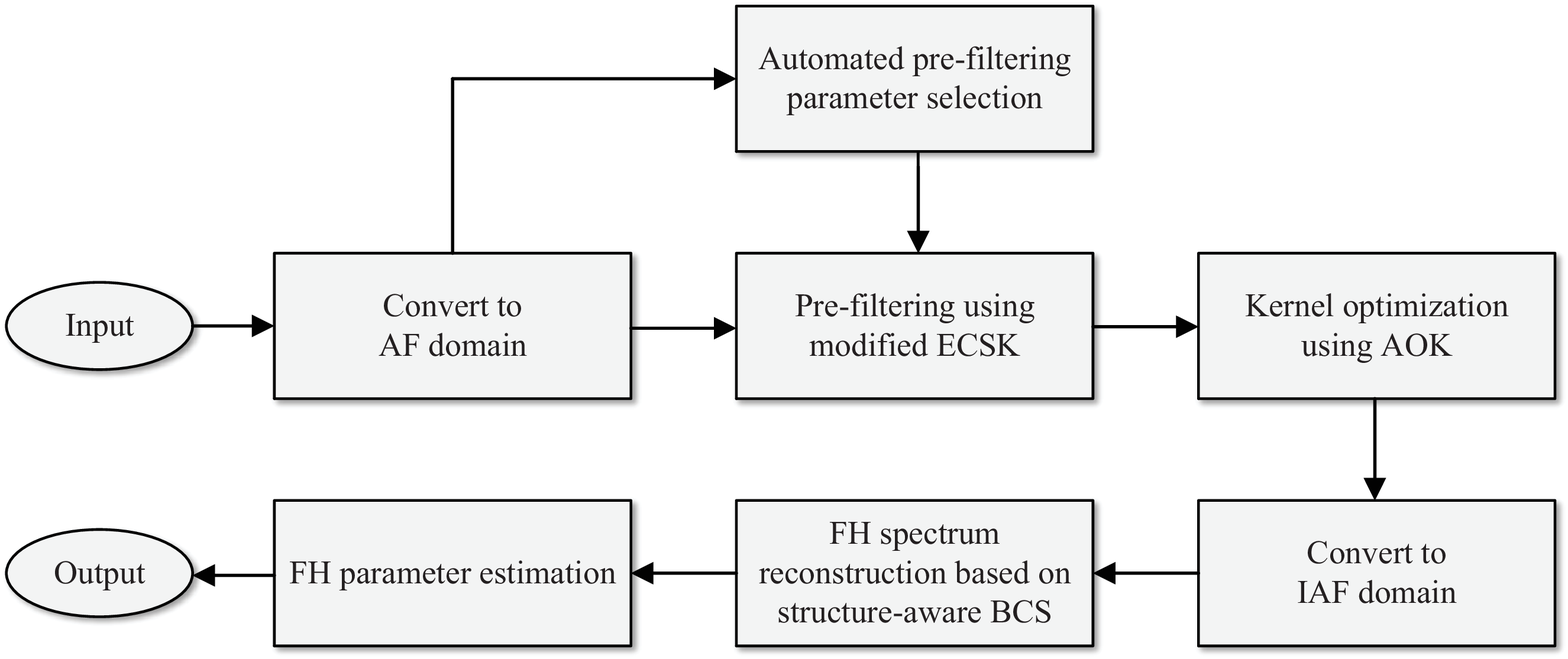}
\caption{Flowchart of the proposed signal processing scheme.}
\label{fig:flow}
\end{figure*}

\subsection{Contributions}

The main novelty of this paper lies in the development of a \emph{Comprehensive Structure-Aware Spectrum Estimation} technique for FH signals which is more advantageous than existing techniques. In particular, it is the first time to investigate the spectrum estimation for FH signals \emph{in the presence of missing observations}. The concept of structure-awareness contains two major components, namely, structure-aware TF kernel design and structure-aware TFR reconstruction. (a) A structure-aware TF kernel is first developed in the AF domain to perform effective suppression of cross-terms and artifacts due to missing observations while preserving the FH signal auto-terms. In particular, we propose a new waveform-adaptive TF kernel design which combines an automatically optimized pre-filtering window and the data-dependent AOK. The pre-filtering window function exploits the prior information of the FH waveform characteristics, whereas the AOK further optimizes the kernel for effective cross-term and artifact reduction while preserving the signal auto-terms. The kernelled AF is then transformed to the instantaneous autocorrelation function (IAF) domain through a one-dimensional (1-D) Fourier transform with respect to (w.r.t.) the frequency difference (Doppler) domain. The IAF results are then processed using sparse reconstruction methods for high-resolution reconstruction of the FH signal TF spectrum. (b) In the sparse reconstruction process, a re-designed BCS approach is developed to estimate the TFD of the signal from the IAF. A novel structure prior for the TFD is imposed to enforce the unique horizontal continuity of the TFR, that characterize the underlying FH signals. Compared with \cite{wu14, yu12}, in addition to designing new structure-aware patterns, we also propose nonlinear updating rules associating the hyper-parameters with the TF patterns, rather than simply select the hyper-parameters from fixed categories. As such, the proposed approach can robustly estimate the FH spectrum in the presence of a high number of missing samples and when the \emph{a priori} information of the hopping patterns is unavailable. As we will show in Section \ref{s:nea}, while existing methods coping with the FH parameter estimation problem developed in \cite{ange10, ange13, zhao15} deteriorate sharply when treating data with missing observations, the proposed approach achieves superior performance in such challenging scenarios.

\medskip

Notations: Lower-case (upper-case) bold characters are used to denote vectors (matrices). ${\rm {abs}}(\cdot)$ returns the modulus of a given complex number. $\circ$ denotes Hadamard product. ${\rm{diag}} \{\cdot\} $ represents a diagonal matrix that uses the entries of a vector as its diagonal entries, and ${\bf I}_N$ denotes an $N\times N$ identity matrix. ${{\bf{F}}_d}$ and ${{\bf{F}}_d^{-1}}$ denote the 1-D discrete Fourier transform (DFT) and inverse discrete Fourier transform (IDFT) matrices w.r.t.\ the $d$ dimension, respectively, and ${{\bf{F}}_{d_1,d_2}}$ denotes a two-dimensional (2-D) DFT w.r.t.\ the $d_1$ and $d_2$ dimensions. $(\cdot)^*$, $(\cdot)^{\rm T}$ and $(\cdot)^{\rm H}$ respectively denote complex conjugate, transpose and Hermitian operations of a matrix. ${\left\| {\bf \cdot}\right\|_p}$ represents the $\ell_p$-norm of a vector, and $\left| {\cdot} \right|$ denotes the cardinality of a set. $p(\cdot)$ denotes the probability density function (PDF). ${\cal B}(\cdot)$, ${\cal {CN}}(\cdot)$, ${\rm {Beta}}(\cdot)$, and ${\rm {Gamma}}(\cdot)$ denote Bernoulli, complex Gaussian, Beta, and Gamma distributions, respectively.

\section{Structure-Aware Adaptive Kernel Design}
\label{s:saakd}

The main stages of the proposed structure-aware scheme are summarized in the flowchart depicted in Fig.\ \ref{fig:flow}. In this section, we first present a detailed description of the proposed signal-dependent kernel design. A joint-variable representation of the FH spectrum in the presence of missing samples is first described in Section \ref{ss:jvr}, and the adaptive kernel design is introduced in Section \ref{ss:akd}. Section \ref{ss:ppo} discusses the optimization of the pre-filtering parameters.

\subsection{Joint-Variable Representations of Missing-Sample FH Spectrum}
\label{ss:jvr}

The discrete-time IAF of signal $x[n]$ is defined as \cite{boas15}
\begin{equation}\label{eq:iaf}
C_{xx}[\tau,n] \triangleq x[n + \tau ]x^ *[n - \tau ],
\end{equation}
where $\tau$ denotes the time-lag index. Stacking $C_{xx}[\tau,n]$ corresponding to all values of $\tau$ and $n$ results in an IAF matrix ${{\bf{C}}_{{\bf{xx}}}}$. Then, the AF matrix of signal vector ${\bf{x}}[n]$, expressed w.r.t.\ lag $\tau$ and Doppler frequency $\kappa$, can be obtained by performing 1-D IDFT on the IAF w.r.t.\ the time index $n$, i.e.,
\begin{equation}\label{eq:af}
{{\bf A_{xx}}}\{ \tau, \kappa \} = {\bf{F}}_n^{ - 1}{{\bf{C}}_{{\bf{xx}}}} \{ \tau, n \}  = \sum\limits_n {{{\bf C_{xx}}} \{ \tau, n \}  {{\rm{e}}^{{\rm{\jmath2\pi }}\kappa n}}},
\end{equation}
where the notation $\{ \tau, \kappa \} $ is used to emphasize that the matrix ${{\bf A_{xx}}}$ is constructed w.r.t.\ variables $\tau$ and $\kappa$. Similarly, the WVD can be obtained by performing 1-D DFT on the IAF w.r.t.\ the lag index $\tau$, i.e.,
\begin{equation}\label{eq:wvd}
{{\bf W_{xx}}} \{ f, n \}  = {{\bf{F}}_\tau }{{\bf{C}}_{{\bf{xx}}}} \{ \tau, n \}  = \sum\limits_\tau  {{{\bf C_{xx}}} \{ \tau, n \} {{\rm{e}}^{{\rm{ - \jmath4\pi }}f\tau}}}.
\end{equation}

\emph{Remarks:} Note that we use ${{\rm{ - \jmath4\pi }}f\tau}$ in the above expression to perform the DFT because integer lags are adopted. This is a common practice in computing the discrete WVD.

\begin{figure*}[!htbp]
\centering
\subfloat[]{\includegraphics[scale=0.67]{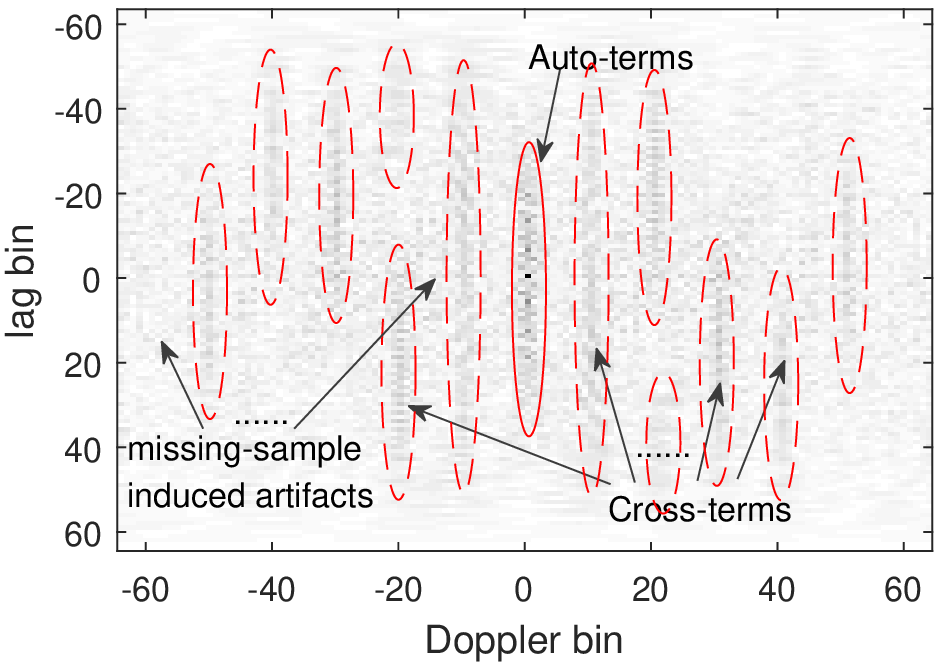}}
\hfil
\subfloat[]{\includegraphics[scale=0.65]{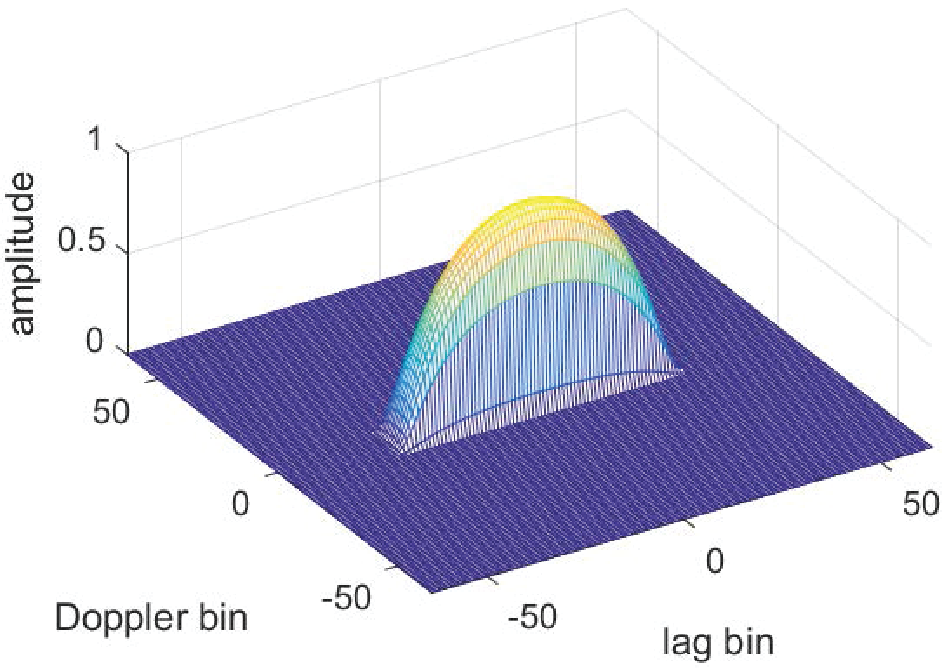}}
\caption{Illustration of undesired terms and kernel design. (a) A typical AF plane of an FH signal. (b) An illustrative example of revised ECSK.}
\label{fig:ker_des}
\end{figure*}

Substituting (\ref{eq:xn}) and (\ref{eq:iaf}) into (\ref{eq:af}), the AF of the observed FH signal with missing samples can be obtained as follows
\begin{equation}\label{eq:af_m}
\begin{array}{l}
{A_{xx}}[\tau ,\kappa ] = \displaystyle\sum\limits_n {{C_{ss}}[\tau ,n]\left( {1 - \sum\limits_{{n_m \in N_m}} {\delta \left[ {n - {n_m} + \tau } \right]} } \right)}\\
 \qquad\qquad\quad\displaystyle\cdot\left( {1 - \sum\limits_{{n_m \in N_m}} {\delta \left[ {n - {n_m} - \tau } \right]} } \right){{\rm{e}}^{\jmath {\rm{2}}\pi \kappa n}}\\
 = {A_{ss}}[\tau ,\kappa ] - \displaystyle\sum\limits_{{n_m \in N_m}} {s[{n_m}]{s^*}[{n_m} - 2\tau ] \cdot {{\rm{e}}^{\jmath {\rm{2}}\pi \kappa (n_m - \tau )}}}\\
  \quad\displaystyle - \sum\limits_{{n_m \in N_m}} {s[{n_m} + 2\tau ]{s^*}[{n_m}] \cdot {{\rm{e}}^{\jmath {\rm{2}}\pi \kappa (n_m + \tau )}}} \\
  \quad\displaystyle  + \displaystyle\sum\limits_n {\sum\limits_{{n_m \in N_m}} {\Bigg(\left( {\delta \left[ {n - {n_m} + \tau } \right]s[n + \tau ]}\right.}}\\
  \qquad\displaystyle{{\left.{\cdot\sum\limits_{{n_l} \ne {n_m}\in N_m} {\delta \left[ {n - {n_l} - \tau } \right]{s^*}[n - \tau ]} } \right)} } {{\rm{e}}^{\jmath {\rm{2}}\pi \kappa n}}\\
  \quad  +\displaystyle\sum\limits_{{n_m\in N_m}} {\delta \left[ { - 2\tau } \right]} s[{n_m}]{s^*}[{n_m} - 2\tau ]{{\rm{e}}^{\jmath {\rm{2}}\pi \kappa (n_m - \tau )}},
\end{array}
\end{equation}
where $C_{ss}$ and $A_{ss}$ respectively denote the IAF and AF of the original FH signal $s[n]$ without missing samples. The term $A_{ss}$ in (\ref{eq:af_m}) contains $\sum\nolimits_{h = 1}^H {h{K_h}}$ auto-terms and ${\left( {\sum\nolimits_{h = 1}^H {h{K_h}} } \right)^2} - \sum\nolimits_{h = 1}^H {h{K_h}}$ cross-terms.

It can be observed from (\ref{eq:af_m}) that, the missing-sample AF consists of two parts, i.e., the full-data AF $A_{ss}$ of $s[t]$ and the artifacts due to missing samples. The latter contains the auto-terms of the missing samples and the cross-terms between the signal and the missing samples. The artifacts expressed in (\ref{eq:af_m}) resemble noise in the sense that they spread over the entire ambiguity domain. The noise pattern of the first two artifact terms in the ambiguity domain depends on the values of the missing observations and their positions, whereas the third artifact term is only affected by the missing-sample positions. As pointed out in references \cite{amin15, Branka15}, careful attention should be paid to the last artifact term, which is always located at $\tau=0$, i.e., along the Doppler frequency axis. This discourages the use of conventional kernels which, due to the required marginal properties, capture all values along the $\tau=0$ axis. A typical AF magnitude plot of an FH signal is depicted in Fig.\ \ref{fig:ker_des}(a).

\subsection{Adaptive Kernel Design}
\label{ss:akd}

With the use of the \emph{a priori} information on the TF structure of the FH signal, i.e., its piecewise constant frequency TF signature, we can apply a proper pre-filtering window before optimizing the AOK so as to prevent the artifacts from being falsely identified as desired signal components and misguiding the AOK optimization process. Generally, for signals whose auto-terms are nearly parallel to either the lag or the Doppler axis, which are exactly the case with the FH signals considered in this paper, the extended compact support kernel (ECSK) outperforms the other kernels in terms of artifact suppression and auto-term preservation \cite{BoasDSP15, boas15, Abed12}. The ECSK also provides flexibility to independently adjust the shape and the size of the kernel. In this paper, we modify the ECSK such that different shape control parameters are used for the two branches, i.e., the lag and Doppler, to offer better flexibility. The modified ECSK is formulated as
\begin{equation}\label{eq:AF_ker_apri}
\tilde g[\tau ,\kappa] = {\tilde g_1}[\tau ] \cdot {\tilde G_2}[\kappa ],
\end{equation}
where
\begin{equation}\label{eq:AF_ka_tau}
{{\tilde g}_1}[\tau] = \left\{ {\begin{array}{*{20}{c}}
{\rm{exp}}\left( {{\rho _1} + \displaystyle\frac{{{\rho _1}\Xi _1^2}}{{{\tau ^2} - \Xi _1^2}}} \right)&{\left| {\tau} \right|  < \Xi_1 N,}\\
{0,}&{{\rm {otherwise,}}}
\end{array}} \right.
\end{equation}
and
\begin{equation}\label{eq:AF_ka_tau}
{{\tilde G}_2}[\kappa] = \left\{ {\begin{array}{*{20}{c}}
{\rm{exp}}\left( {{\rho _2} + \displaystyle\frac{{{\rho _2}\Xi _2^2}}{{{\kappa ^2} - \Xi _2^2}}} \right)&{\left| {\kappa} \right|  < \Xi_2 N,}\\
{0,}&{{\rm {otherwise,}}}
\end{array}} \right.
\end{equation}
respectively represent lag/Doppler window branches. In the above expressions, $\rho_1$ and $\rho_2$ denote the shape control parameters of the two branches, and $\Xi_1$ and $\Xi_2$ represent their respective sizes. Larger values of $\rho_1$ and $\rho_2$ result in a steeper kernel shape in the corresponding branch, whereas larger values of $\Xi_1$ and $\Xi_2$ imply a larger kernel size.

In our proposed method, prior to the radial kernel optimization procedure, the short-time AF is pre-filtered by utilizing the modified ECSK \emph{in a time-localized, short-time manner}, as illustrated in Fig.\ \ref{fig:ker_des}(b), where the preserved support for the auto-terms is a sufficiently small region where the Doppler frequency is nearly zero. In doing so, the vertical TFD stripes due to impulsive missing samples, whose AF components spread in the Doppler domain, and noise-like artifacts, whose AF components spread in the entire ambiguity domain, are effectively eliminated.

\subsection{Pre-filtering Parameter Optimization}
\label{ss:ppo}

Enhanced TFD concentration generally yields sharp TF representations and reduced vicinal interference. To achieve an optimal pre-filtering performance that simultaneously maximizes the TFD concentration and minimizes the TFD artifacts, parameter pairs $\rho_1$ and $\rho_2$ as well as $\Xi_1$ and $\Xi_2$ should be tuned to their optima based on a proper criterion. Several optimization criteria are available in the literature for the evaluation of the concentration performance. Among these criteria, distribution norm-based measures \cite{jones90, jones94} and entropy-based measures \cite{Baraniuk01, Aviyente05} are commonly used. However, norm-based measures tend to discriminate poorly concentrated components \cite{stan01}, whereas entropy-based measures are sensitive to amplitude and phase variations \cite{Baraniuk94}. In this context, an efficient energy concentration measure is introduced in \cite{stan01} that overcomes the aforementioned drawbacks, and has been applied to automatic determination of the best window length in the computation of spectrogram. In this paper, an automatic parameter tuning method is proposed based on this energy concentration measure. The discrete-time expression of this energy concentration measure can be written as
\begin{equation}\label{eq:con_meas}
{\cal M}\left( {{{\bf{F}}_{\tau,\kappa}}{{\widetilde {\bf{A}}}_{{\bf{xx}}}}\{ n;\tau ,\kappa \}} \right)={\left( {\sum\limits_{\tau} {\sum\limits_{\kappa}{{{\left| {{{\bf{F}}_{\tau,\kappa}}{{\widetilde {\bf{A}}}_{{\bf{xx}}}}\{ n;\tau ,\kappa \}} \right|}^2}} } } \right)^2}.
\end{equation}
We define the cost function in our pre-filtering parameter optimization process as
 \begin{equation}\label{eq:cost_op}
\begin{array}{l}
\mathop {\min }\limits_{{\rho _1},{\rho _2},{\Xi _1},{\Xi _2}} {\cal M}\left( {{{\bf{F}}_{\tau,\kappa}}{{\widetilde {\bf{A}}}_{{\bf{xx}}}}\{ n;\tau ,\kappa \}}; {{\rho _1},{\rho _2},{\Xi _1},{\Xi _2}} \right) \\
\begin{array}{*{20}{l}}
	{{{\rm{s.t.}}}\quad 0.01 \le {\rho _1},{\rho _2} \le 10,}\\
	{\quad\quad\;0.01 \le {\Xi _1},{\Xi _2} \le 0.5.}
\end{array}
\end{array}
\end{equation}
The constraints in (\ref{eq:cost_op}) are set according to the domain of definition and can are applicable to different types of FH signals.

\begin{figure}[!htbp]
\centering
\includegraphics[scale=0.66]{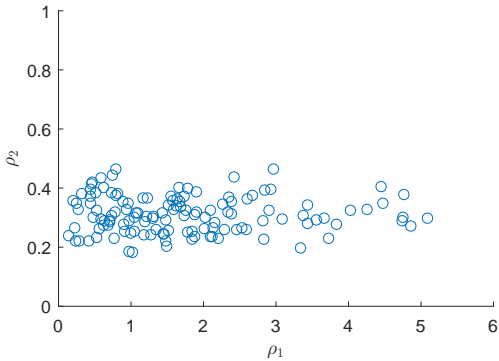}
\caption{Illustrative scatter plot of optimized parameters.}
\label{fig:patterns}
\end{figure}

To achieve a fully automated optimization of the kernel parameters, an adaptive differential evolution algorithm \cite{zhangj09} is adopted. When the \emph{a priori} knowledge about the distribution of potential optima is available, we can further exploit it to arrange the initial population settings. In the simulation examples provided in Section \ref{s:nea}, we assume that the potential optima follow a uniform distribution. To better illustrate the parameter optimization process, we provide a $128$-point scatter plot of the optimized parameters $\rho_1$ and $\rho_2$ in our numerical trials. It can be observed that the optimal values of $\rho_2$ vary within a relatively narrower range than $\rho_1$.

\subsection{AOK After Pre-filtering}
\label{ss:ppo}

After applying the modified ECSK as a pre-filtering window, AOK is then employed to further mitigate the effect of artifacts due to missing samples. As discussed in Section \ref{ss:jvr}, such artifacts spread over the entire ambiguity domain. The AOK is a well-known data-dependent kernel, which is designed by solving the following optimization problem \cite{jone95}:
\begin{equation}\label{eq:aok}
\begin{array}{l}
{\Phi _{{\rm{opt}}}}(r,\psi ) = \mathop {\arg \max }\limits_{\Phi (r,\psi )} \displaystyle \int_0^{2\pi } {\int_0^\infty  {{{\left| {A(r,\psi )\Phi (r,\psi )} \right|}^2}r{\rm{d}}r{\rm{d}}\psi } } \\
\begin{array}{*{20}{l}}
	{{{\rm{s.t.}}}\quad\displaystyle\Phi (r,\psi ) = \exp \left( { - \frac{{{r^2}}}{{2{\sigma ^2}(\psi )}}} \right),}\\
	{\quad\quad\;\displaystyle\frac{1}{{4{\pi ^2}}}\int_0^{2\pi } {{\sigma ^2}(\psi )d\psi }  \le \alpha ,}
\end{array}
\end{array}
\end{equation}
where $\alpha$ denotes the kernel volume constraint, $A(r,\psi )$ represents the AF of the signal in polar coordinates, and $r$ and $\psi$ denote the radius and radial angle variables, respectively. Equation (\ref{eq:aok}) is optimized in the sense that the signal auto-terms are preserved to the maximum extent within the low-pass Gaussian filter, while the pass-band area of the filter is limited to a total volume of $\alpha$ so as to filter out the cross-terms which are located away from the origin, and to reduce the artifacts and noise that spread over the entire ambiguity domain. The desired resolution and the cross-term attenuation are determined by a proper selection of $\alpha$.

For signals with time-varying characteristics, AOK is usually implemented with a time-localized short-time AF. At time instant $t$, a time-adaptive kernel $\Phi _{\rm{opt}}(t;r,\psi)$ is produced by substituting the short-time AF $A(t;r,\psi)$ for $A(r,\psi)$ in (\ref{eq:aok}) and then following the polar-coordinate Gaussian kernel optimization procedure for each individual TFD slice \cite{jone95}. Denoting the rectangular-coordinate short-time AF as $A[n;\tau ,\kappa ]$, the pre-filtered short-time AF can be expressed as
\begin{equation}\label{eq:prewin}
\begin{array}{l}
\displaystyle\tilde A_{xx}[n;\tau ,\kappa] = \tilde g[n;\tau ,\kappa ] \cdot A_{xx}[ {n;\tau ,\kappa } ]\\
\qquad\qquad= \displaystyle\tilde g[n;\tau ,\kappa ] \cdot \int {x[u + \tau ]w[u - n + \tau ]} \\
\quad\qquad\qquad\displaystyle\cdot {x^*}[u - \tau ]{w^*}[u - n - \tau] {{\rm{e}}^{{\rm{\jmath}}\kappa u}}{\rm{d}}u,
\end{array}
\end{equation}
where $w[n]$ represents a rectangular short-time sliding window. Stacking $\tilde A_{xx}[n;\tau ,\kappa ]$ for all $\tau$ and $\kappa$ results in the short-time AF matrix ${{\widetilde {\bf{A}}}_{{\bf{xx}}}}\{ n;\tau ,\kappa \}$. Then, the TFD corresponding to the kernelled AF is obtained as its 2-D DFT w.r.t.\ $\tau$ and $\kappa$, expressed as
\begin{equation}\label{eq:tfraok}
\displaystyle{{{\bf{\tilde W}}}_{{\bf{xx}}}}\{ f,n\}  = {{\bf{F}}_{\tau,\kappa}}{{\widetilde {\bf{A}}}_{{\bf{xx}}}}\{ n;\tau ,\kappa \} {{\bf \Phi} _{{\rm{opt}}}}\{n;\tau ,\kappa\},
\end{equation}
where ${\bf \Phi}\{n;\tau, \kappa\}$ is the time-localized AOK matrix represented in the rectangular $(\tau, \kappa)$ coordinate system.

\emph{Remarks:} It is worth emphasizing that, when compared to references \cite{ange10, ange13, zhao15}, which consider FH spectrum estimation in the context of linear short-time Fourier transform (STFT), the utilization of the bilinear TFR in this paper enables us to better address the missing-sample problem because kernel design and its capability to filter out undesired signal components can be utilized \emph{only} in bilinear TF analysis. This is a key novel contribution of this paper since so far only the linear TF analysis has been used in sparse FH spectrum estimation, and no missing samples have been considered in the literature.

\begin{figure*}[!b]
\centering
\includegraphics[scale=0.4]{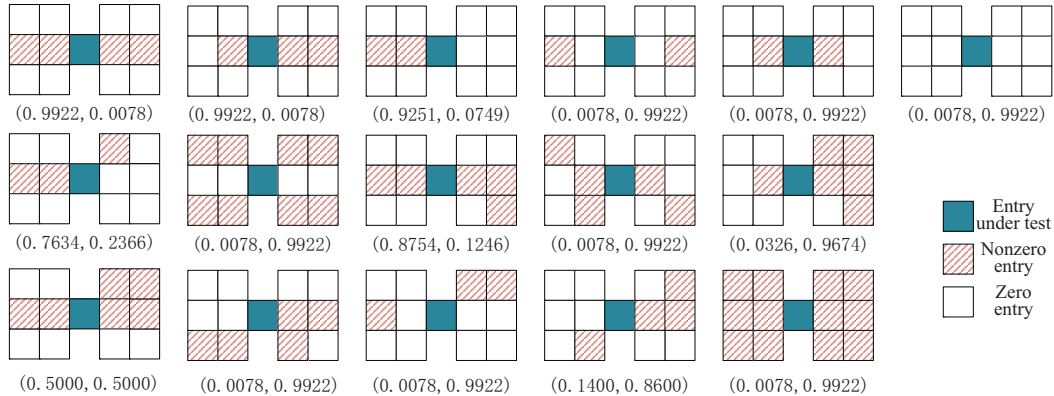}
\caption{Examples of different $3\times 5$ TF patterns.}
\label{fig:patterns}
\end{figure*}

\section{BCS-Based FH Spectrum Estimation}
\label{s:sfjs}

\subsection{CS Model for FH Spectrum Estimation}

In this section, we consider a CS based approach which yields a high-resolution TFR. The IAF matrix corresponding to the kernelled AF is obtained as the 1-D IDFT of ${\tilde {\bf A}}_{\bf xx}\{ \tau, \kappa \}$ w.r.t.\ $\kappa$, i.e.,
\begin{equation}
{\tilde {\bf C}}_{\bf xx} \{ \tau, n \} = {\bf F}^{-1}_{\kappa} {\tilde {\bf A}}_{\bf xx}\{ \tau, \kappa \}.
\end{equation}
On the other hand, the bilinear TFR matrix is associated with the IAF matrix by the following 1-D Fourier relationship:
\begin{equation}
{\tilde {\bf C}}_{\bf xx} \{ \tau, n \} = {\bf{F}}_f^{ - 1} {{\bf{\tilde W}}_{{\bf{xx}}}} \{ f, n \}.
\end{equation}

The CS approach obtains ${{\bf{\tilde W}}_{{\bf{xx}}}} \{ f, n \}$ by exploiting the above Fourier transform relationship but through a sparse reconstruction operation. Denote ${\tilde {\bf c}}_{\bf xx}[n]$ as the $n$-th column of the IAF matrix ${{\tilde {\bf C}}_{\bf xx}} \{ \tau,n \}$, and ${\tilde {\bf{w}}_{{\bf{xx}}}} [n]$ as the $n$-th column of the bilinear TFR matrix ${\bf{\tilde W}}_{\bf xx} \{ f, n \}$. Then, their relationship conforms to the following standard linear model commonly used in CS and sparse reconstruction:
\begin{equation}\label{eq:CS_FH_model}
{\tilde {\bf c}}_{\bf xx}[n] = {\bf{F}}_f^{ - 1} {\tilde {\bf{w}}_{\bf{xx}}} [n].
\end{equation}
Therefore, the TFR can be obtained from sparse reconstruction, in lieu of conventional Fourier transform, by repeating the procedure for each time instant. Various CS algorithms can be used for this purpose. In the following, we consider this problem from a BCS perspective \cite{ji08}, and the structure of the FH spectrum is utilized for improved spectrum estimation. BCS methods are known for their capability to flexibly model sparse signals that not only promote the sparsity of its solution, but also exploit additionally known structures of the sparse signal \cite{zai13}. For notational convenience, we simplify the notations ${\tilde {\bf c}}_{\bf xx}[n]$, ${\bf{F}}_f^{ - 1}$ and ${\tilde {\bf{w}}_{\bf{xx}}} [n]$ as ${\bf c}$, ${\boldsymbol \Lambda}$ and ${\bf w}$, respectively, i.e.,
\begin{equation}\label{eq:CS_FH_model_simple}
{\bf c} = {\boldsymbol \Lambda} {\bf w}.
\end{equation}

\subsection{Sparsity Prior}

\begin{figure*}[!b]
\normalsize
\hrulefill
\newcounter{MYtempeqncnt}
\setcounter{MYtempeqncnt}{\value{equation}}
\setcounter{equation}{25}
\begin{equation}\label{eq:stru_fun}
{z_{{\rm{ver}}}},{z_{{\rm{hor}}}} \to \varpi :\;\varpi  \triangleq \left( {1 - {{\left( {\frac{1}{2}} \right)}^{{{\left( {\frac{1}{2}\left( {\sqrt {\left( {1 + 4\times 2} \right)\frac{{1 + 2{z_{{\rm{ver}}}}}}{{1 + {z_{{\rm{hor}}}}}}}  - 1} \right)} \right)}^4}}}} \right) + {\left( {\frac{1}{2}} \right)^{z_{{\rm{hor}}}^2}},
\end{equation}
with
\begin{equation}\label{eq:z_ver}
{z_{\rm{ver}}} \triangleq \frac{1}{2}\left( {{z_{{{\cal J}_{ \otimes i + }}}} + {z_{{{\cal J}_{ \otimes i - }}}} + \sum\limits_{j = 1}^2 {{z_{i + j}}\left( {{z_{\left( {i + j} \right) + }} + {z_{\left( {i + j} \right) - }}} \right)} } \right),
\end{equation}
\begin{equation}\label{eq:z_hor}
{z_{{\rm{hor}}}} \triangleq \left\{
{\begin{array}{*{30}{l}}
0,&{\prod\limits_{j = 1}^2 {{z_{i + j}} + \prod\limits_{j = 1}^2 {{z_{i - j}}}  = 0},}\\
\displaystyle\sum\limits_{j = 1}^2 \Big({\left( {{z_{i + 1}} + {z_{i - 1}}} \right){\rm{ + }}\left( {{z_{i + 1}}{z_{i + 2}} + {z_{i - 1}}{z_{i - 2}}} \right){\rm{ + }} \ldots } &\\
\qquad\qquad+\left( {{z_{i + 1}}{z_{i + 2}} \ldots {z_{i + j}} + {z_{i - 1}}{z_{i - 2}} \ldots {z_{i - j}}} \right)\Big),\qquad&{\rm{otherwise.}}
\end{array}} \right.
\end{equation}
\setcounter{equation}{\value{MYtempeqncnt}}
\end{figure*}

The BCS is a nonparametric solver of sparse linear inverse problems imposing a conditional Gaussian prior with its precision (reciprocal of the variance) guided by a hyperprior of Gamma distribution, i.e., $\alpha_0 \sim {\mathop{\rm Gamma}\nolimits} (c,1/d)$. The BCS assumes the following likelihood model \cite{wipf04}

\begin{equation}\label{eq:lh_input}
p\left({\bf{c}}; {{\bf w},{\gamma _0}}\right)  = {\cal {CN}}({\bf{c}};{{\boldsymbol \Lambda}}{\bf {w}},{\gamma _0}{\bf{I}}),
\end{equation}
where $\gamma _0=\alpha_0^{-1}$ is the variance. To encourage sparsity of the FH signal TFR, a Dirichlet process prior with a spike-and-slab centering distribution \cite{wu14, yu11} is employed to $w_i$, i.e., the $i$-th entry of ${\bf w}$, which allows different predictors to have identical coefficients while performing variable selection. That is,
\begin{equation}\label{eq:sspri}
p({w _i};{\gamma _i},{\pi _i}) = (1 - {\pi _i})\delta_0 + {\pi _i}{\cal {CN}}(w _i;0,\gamma _i),
\end{equation}
where $\pi _i$ is a mixing weight standing for the prior probability of a nonzero entry, and $\delta_0$ represents the delta function with a unit point measure concentrated at zero. Also, we assign a Gamma prior to the precision as $\gamma _i^{ - 1}=\alpha _i \sim {\mathop{\rm Gamma}\nolimits} (a,1/b)$.

To make the inference analytical, a product of two latent variables $z_i$ and $\theta _i$, i.e., $w _i=z_i\cdot\theta _i$, is introduced to follow the PDF in (\ref{eq:sspri}), where $\theta _i \sim {\cal {CN}}(\theta _i;0,\gamma _i)$, and $z_i$ is a binary variable that follows the
Bernoulli distribution ${\cal {B}}(\pi _i)$. $z_i = 1$ implies that the $i$-th entry is nonzero, whereas $z_i=0$ implies a zero entry. Denote ${\bf z}=\left[z_1, \ldots, z_N\right]^{\rm T}$ and ${\boldsymbol \theta}=\left[\theta_1, \ldots, \theta_N\right]^{\rm T}$. The overall prior on ${\boldsymbol{\theta}}$ w.r.t. $a$ and $b$ can be evaluated analytically through the integration over ${\boldsymbol{\alpha}}$, and it corresponds to the Student-t distribution \cite{tipp01}. With an appropriate choice of $a$ and $b$, the Student-t distribution is strongly peaked about ${\boldsymbol{\theta}}=0$, and thus the overall prior on ${\boldsymbol{\theta}}$ favors sparseness \cite{yu11}. In practice, the hyper-parameters $a$, $b$, $c$, $d$ are usually assigned to small values to make the corresponding priors flat.

\subsection{Structure Prior}

The FH spectrum shows sparse piecewise constant frequencies. This structure characteristic can be exploited to improve the accuracy and robustness of the sparse learning performance. For the underlying FH signals, a continuous structure prior that encourages the FH spectrum to have a longer horizontally linear structure in the TFR is desired. With a slightly increased computational complexity, we extend the model to size $3 \times 5$, i.e., the neighborhood entries that are within a Euclidean distance of $2$, and the vertical pixels from the proximate frequency rows are taken into consideration when decision is made to the TF entry under test. It is evident from Fig.\ \ref{fig:patterns} that the utilization of the structure model with a higher dimension enables more comprehensive characterization and treatment of pixel patterns. In this case, simply dividing various patterns into a fixed number of categories does not adequately characterize the relationship between the neighboring pixels. In addition, in the situations with a low SNR, there will be more artifact residue and a higher level of spectrum distortion. As a result, simply rejecting all entries with vertical non-zero neighbors will degrade the robustness of the algorithm.

\begin{figure}[!tphb]
\centering
\subfloat[]{\includegraphics[scale=0.6]{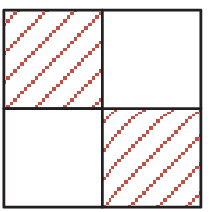}}
\hfil
\subfloat[]{\includegraphics[scale=0.6]{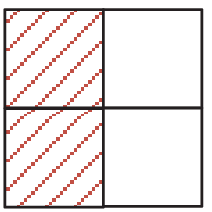}}
\caption{Two patterns of vertical adjacent structure. (a) Indirectly adjacent. (b) Directly adjacent.}
\label{fig:ver_patt}
\end{figure}


In this context, we propose a new structure prior which is related to each individual pattern with a proper nonlinear relationship. We first define the neighborhood of index $i$ as
\begin{equation}\label{eq:neigb}
{\cal J}_{ \odot i} \triangleq \left\{ {j\left| {d\left( {i,j} \right) \le 2,j \in [1,N]} \right.} \right\},
\end{equation}
where $d\left( {i,j} \right)$ is the Euclidean distance between $i$ and $j$. We then define the deleted neighborhood of index $i$, i.e., the neighborhood of index $i$ with $i$ itself excluded, as
\begin{equation}\label{eq:dele_neigb}
{\cal J}_{ \otimes i} \triangleq \left\{ {j\left| {d\left( {i,j} \right) \le 2,j \in [1,N]} \right.,j \ne i} \right\}.
\end{equation}
The number of nonzero entries at a location $i$ and its neighborhood is denoted as ${z_{{\cal J}_{ \odot i}}}$. Note that during the pattern classification process, three rows of ${\bf z}$ are investigated, and we denote the location $i$ in the $[n+1]$-th and $[n-1]$-th rows of ${\bf z}$ as $i+$ and $i-$, respectively. In Bayesian probability theory, if the posterior distributions belong to the same family as the prior probability distribution, then the prior and posterior are termed conjugate distributions. The Beta distribution is conjugate to Bernoulli likelihood, so $\pi_i$ is assumed to follow the Beta distribution. For a certain structure prior, the posterior distribution of $\pi _i$ is derived as
\begin{equation}\label{eq:mix_pattern}
p(\pi _i;{e},{f})
 = {\rm{Beta}}\left( {{e} + {{z_{{\cal J}_{ \odot i}}}},{f} + \left| {{\cal J}_{ \odot i}} \right| - {{z_{{\cal J}_{ \odot i}}}}} \right).
\end{equation}

The ${\rm {Beta}}(e,f)$ distribution tends to draw small values of $\pi_i$ when $e<f$, and a large value when $e>f$, while it has no tendency when $e=f$. By choosing proper values of $e$ and $f$, therefore, we can encourage or discourage the sparsity of the pixel under test, depending on the sparsity support in the neighboring pixels. The value of hyper-parameters $e$ and $f$ should be decimal fractions between $0$ and $1$. In the previous three-decision-category based method \cite{wu14, yu12}, these hyper-parameters are multiples of $1/N$, and $N$ is chosen to be integer power of $2$ for computation efficiency. Also, as a rational nonlinear relationship associating the hyper-parameters with the TF structure should encourage longer horizontal structures while discourage high vertical-to-horizontal non-zero neighborhood ratios, we choose a straightforward formula $\left( {1 - {{\left( {\frac{1}{2}} \right)}^{{\rm {func}}\left( {{z_{{\rm{ver}}}}/{z_{{\rm{hor}}}}} \right)}}} \right) + {\left( {\frac{1}{2}} \right)^{{\rm {func}}\left( {{z_{{\rm{hor}}}}} \right)}}$ to express the nonlinear relationship between hyper-parameter $f$ and the TF structure patterns, and let $e=1-f$. On the other hand, some modifications should be made to the formula to ensure that the value of $f$ corresponds to the boundary values of ${z_{{\rm{hor}}}}$, whereas ${z_{{\rm{ver}}}}$ is constrained to a reasonable range and to avoid zero denominator. As a result, the hyper-parameter $f$ can be derived as

\begin{figure*}[!b]
\normalsize
\hrulefill
\setcounter{MYtempeqncnt}{\value{equation}}
\setcounter{equation}{38}
\begin{equation}\label{eq:intewc}
\begin{array}{l}
p\left( {{\bf{w }}\left| {\bf{c}} \right.} \right)\propto \int {p\left( {{\bf w} \left|{\boldsymbol {\gamma }} ,{\boldsymbol {\pi }}, {\bf c},{\alpha _0} \right.} \right){\rm d}{\boldsymbol {\gamma }}{\rm d}{\boldsymbol {\pi }}{\rm d}{\alpha _0}}\\
\qquad\quad\;\;\propto \displaystyle{\left( {d + \frac{1}{2}\left\| {{\bf{c}} - {\bf{\Lambda w}}} \right\|_2^2} \right)^{ - c - \frac{N}{2}}}\prod\limits_{i = 1}^N \left({\frac{{\Gamma \left( {a{\rm{ + }}\frac{{{z_{{{\cal J}_{ \odot i}}}}}}{2}} \right)}}{{{{\left( {b{\rm{ + }}\frac{{\left\| {{z_{{{\cal J}_{ \odot i}}}}} \right\|_2^2}}{2}} \right)}^{a{\rm{ + }}\frac{{{z_{{{\cal J}_{ \odot i}}}}}}{2}}}}} \cdot {\rm{Beta}}\left( {e + {z_{{{\cal J}_{ \odot i}}}},f + \left| {{{\cal J}_{ \odot i}}} \right| - {z_{{{\cal J}_{ \odot i}}}}} \right)}\right).
\end{array}
\end{equation}
\setcounter{equation}{\value{MYtempeqncnt}}
\end{figure*}

\begin{equation}\label{eq:revise_f}
f = \left\{ {\begin{array}{*{20}{l}}
{1/N,}&{\varpi  < 1/N,}\\
\varpi, &{1/N \le \varpi  < 1,}\\
{1 - 1/N,}&{\varpi  \ge 1,}
\end{array}} \right.
\end{equation}
where the value of $\varpi$ is determined in (\ref{eq:stru_fun})-(\ref{eq:z_hor}).

\emph{Remarks:} The statistical properties of Beta distribution, such as mode, mean, and variance are closely related to the weight of each shape parameters in their summation. We set these parameters $e$ and $f$ in order to encode the structure prior beliefs. For example, the mean of the Beta distribution in this paper is set to $({1-\varpi+ {{z_{{\cal J}_{ \odot i}}}}) {\left/ \right.} {\left| {{J_{ \odot i}}} \right|}}$. A similar parameter setting has also been adopted in several existing references (c.f., e.g., \cite{wu14, yu12}). We set $e=1-f$ and keep both hyperparameters for better interpretation and consistency with the existing references.

In the above expressions, the vertical structure factor ${z_{\rm{ver}}}$ is assigned different weights $1/2$ and $1$ respectively to indirectly and directly adjacent structures as shown in Fig.\ \ref{fig:ver_patt}. The reason we discriminate between indirectly and directly adjacent structures is that directly adjacent structures tend to broaden the signal bandwidth. This is contradictory to the fact that the underlying FH signals are instantaneously narrowband. In contrast, indirectly adjacent structures may be formed by the distortion of the desired signal component, noise, or artifact residue, so the weight should be relatively smaller. On the other hand, the horizontal structure factor ${z_{\rm{hor}}}$ is obtained by counting the number of continuous adjacent entries. Note that, if the entry under test is located in an isolated line, i.e., both the left and right edge pixels are $0$, the value of ${z_{\rm{hor}}}$ in (\ref{eq:z_hor}) will be set to 0. Because the codomain of ${z_{\rm{ver}}}$ and ${z_{{\rm{hor}}}}$ can be derived as $\left[ {0,4} \right]$ and $\left[ {0,8} \right]$, respectively, according to (\ref{eq:z_ver}) and (\ref{eq:z_hor}), we can further obtain the domain of $\varpi$ as $\varpi  \in \left[ {{{\left( {1/2} \right)}^{16}},2 - {{\left( {1/2} \right)}^{256}}} \right]$.

Assume that $N=128$. According to the proposed structure prior formation method, the hyper-parameter pairs $\left( {e,f} \right)$ for all the patterns are listed under each case in Fig.\ \ref{fig:patterns}. These hyper-parameters better reflect the corresponding cluttering situation, by automatically assigning a moderate value to the pattern where a long straight line is present whereas the vertical pixels in the nearby rows take a small value. For those cases where nonzero entries extend in the frequency domain or occur isolatedly, a discouraging value will be asserted to prevent or restrain the structure.

\subsection{Bayesian Inference}

Since no closed-form expressions of the Bayesian estimators can be derived, Markov-chain Monte Carlo sampling is used to implement the inference. The maximum likelihood estimation of $w_i$ and $\gamma_i$ from (\ref{eq:sspri}) will generally lead to severe overfitting. To obviate the overfitting problem, a smoother inference model is formulated by defining an automatic relevance determination Gaussian prior over the weights \cite{tipp01}:
\setcounter{equation}{28}
\begin{equation}\label{eq:wgt_pri}
p\left( {{\bf w}; {\boldsymbol {\gamma }} ,{\boldsymbol {\pi }}} \right) = \prod\limits_{i = 1}^N \left[ { (1 - {\pi _i})\delta_0 + {\pi _i} {\cal {CN}}\left( {{w_i}; {0,{\gamma _i}}} \right) } \right],
\end{equation}
where ${\boldsymbol {\gamma }}={\left[ {{\gamma _1}, \ldots ,{\gamma _N}} \right]^{\rm{T}}}$ is a vector consisting of $N$ hyper-parameters that independently control the prior variance of each weight. We can then acquire the posterior distribution of ${\bf w}$ by combining (\ref{eq:wgt_pri}) with the observation likelihood $p\left( {{\bf c}; {{\bf w},{\gamma _0}} } \right)$ in (\ref{eq:lh_input}), i.e.,
\begin{equation}\label{eq:wgammapi}
p\left( {{\bf w} \left|{\boldsymbol {\gamma }} ,{\boldsymbol {\pi }}, {\bf c},{\alpha _0} \right.} \right) \propto p\left( {{\bf w} ;{\boldsymbol {\gamma }} ,{\boldsymbol {\pi }}} \right)p\left( {{\bf c}; {{\bf w},{\alpha _0}}} \right).
\end{equation}

A Gibbs sampler is adopted to implement the Bayesian inference as following. Let ${\boldsymbol \lambda} _i$ be the $i$-th column of ${\boldsymbol \Lambda}$. Then, the paired Gibbs sampler iteratively samples the observations from the following conditional PDF \cite{wu14, yu11}
\begin{equation}\label{eq:gibbs}
\begin{array}{l}
p\left( {{w_i}\left| {{{\bf w}_{\backslash i}},{\bf{c}}} \right.} \right) = p\left( {{\theta _i},{z_i} \left| {{{\boldsymbol{\theta }}_{\backslash i}},{{\bf{z}}_{\backslash i}},{\bf{c}}} \right.} \right)\\
\qquad\qquad= p\left( {{\theta _i}\left| {{z_i},{{\boldsymbol{\theta }}_{\backslash i}},{{\bf{z}}_{\backslash i}},{\bf{c}}} \right.} \right)p\left( {{z_i}\left| {{{\boldsymbol{\theta }}_{\backslash i}},{{\bf{z}}_{\backslash i}},{\bf{c}}} \right.} \right),
\end{array}
\end{equation}
where the notation $(\cdot)_{\backslash i}$ denotes the subvector excluding the $i$-th entry. The probability $p\left( {{z_i}=1\left| {{{\boldsymbol{\theta }}_{\backslash i}},{{\bf{z}}_{\backslash i}},{\bf{c}}} \right.} \right)$ is acquired as
\begin{equation}\label{eq:zi1theta}
p\left( {{z_i}=1\left| {{{\boldsymbol{\theta }}_{\backslash i}},{{\bf{z}}_{\backslash i}},{\bf{c}}} \right.} \right)=\frac{{{\alpha _i}}}{{1 - {\alpha _i}}}\frac{{{\cal {CN}}\left( {0,{\gamma _i}} \right)}}{{{\cal {CN}}\left( {{{\tilde \mu }_i},{\gamma _i}} \right)}},
\end{equation}
where ${\tilde \mu }_i$ and ${{\tilde \gamma }_i}$ are respectively updated as
\begin{equation}\label{eq:til_mu}
{{\tilde \mu }_i} = {\alpha _i^{ - 1}}\alpha _0 {\boldsymbol \lambda} _i^H{{\bf c}_{\backslash i}},
\end{equation}
\begin{equation}\label{eq:til_gam}
{{\tilde \gamma }_i^{- 1}} = {\tilde \alpha }_i= {\left( {\alpha _0 {\boldsymbol \lambda} _i^H{{\boldsymbol \lambda} _i} + \alpha _i} \right)}.
\end{equation}

The conditional distribution of $p\left( {{\theta _i}\left| {{z_i}=1,{{\boldsymbol{\theta }}_{\backslash i}},{{\bf{z}}_{\backslash i}},{\bf{c}}} \right.} \right)$ can be expressed as
\begin{equation}\label{eq:gibbs_thetai}
p\left( {{\theta _i}\left| {{z_i}=1,{{\boldsymbol{\theta }}_{\backslash i}},{{\bf{z}}_{\backslash i}},{\bf{c}}} \right.} \right)={\cal {CN}}\left( {{w_i};{{\tilde \mu }_i},{\gamma _i}} \right).
\end{equation}
For the ${z_i}=0$ case, as the value of ${\theta _i}$ does not affect the result of $w_i$, we directly draw the value of ${\theta _i}$ from its prior. Subsequently, the Gibbs sampler updates the mixing weight $\pi_i$ according to (\ref{eq:mix_pattern}).

Next, we update the precision variable $\alpha_i$. By utilizing the conjugate property of the Gaussian and Gamma distributions, we analytically acquire the posterior distribution of $\alpha _i$ as
\begin{equation}\label{eq:alpha_update}
p\left( {{\alpha _i};a,b,{z _{{\cal J}_{ \odot i}}}} \right) = {\rm{Gamma}}\left( {a{\rm{ + }}\frac{{{z_{{{\cal J}_{ \odot i}}}}}}{2},\frac{1}{{b{\rm{ + }}\frac{{\left\| {{z_{{{\cal J}_{ \odot i}}}}} \right\|_2^2}}{2}}}} \right).
\end{equation}
After completing all the $i$ iterations, the posterior distribution of the noise precision ${\alpha _0}$ is updated as
\begin{equation}\label{eq:alpha0_update}
\begin{array}{l}
p\left( {{\alpha _0};c,d,{\bf{c}},{\boldsymbol \Lambda},{\bf{\theta }},{\bf{z}}} \right)\\
\quad=\displaystyle{\rm{Gamma}}\left( {c{\rm{ + }}\frac{{{\rm{rank}}\{ {\bf{\Lambda }}\} }}{2},\frac{1}{{d{\rm{ + }}\frac{{\left\| {{\bf{c}} - {\bf{\Lambda }}\left( {{\bf{\theta }}\circ {\bf{z}}} \right)} \right\|_2^2}}{2}}}} \right).
\end{array}
\end{equation}

The maximum a posteriori (MAP) estimator is adopted to infer the estimation of ${\bf w}$ as
\begin{equation}\label{eq:esti_theta}
{\bf{\hat w }} =  \arg \mathop {\max }\limits_{\bf{w }} p\left( {{\bf{w }}\left| {\bf{c}} \right.} \right),
\end{equation}
where marginal distribution $p\left( {{\bf{w }}\left| {\bf{c}} \right.} \right)$ can be obtained by integrating out the hyper-parameters ${\boldsymbol {\gamma }}$, ${\boldsymbol {\pi }}$, and ${\alpha _0}$ in (\ref{eq:wgammapi}), as expressed in (\ref{eq:intewc}) \cite{yu11}, where $\Gamma \left( u \right) \triangleq \int_0^\infty  {{t^{u - 1}}{{\rm{e}}^{ - t}}{\rm{d}}t}$ denotes a Gamma function.

This completes the sparse reconstruction result of (\ref{eq:CS_FH_model}) for one time instant. The estimation of the entire FH spectrum is rendered by repeating the BCS-based estimation for each column of ${\bf{\tilde W}}_{\bf xx} \{ f, n \}$.

\emph{Remarks:} The proposed method in this paper differs from that of \cite{liu16c2} in two aspects: (a) In \cite{liu16c2} a threshold-based post-AOK window was adopted, whereas in this paper we pre-filter the running AF with a new ECSK. ECSK is known as the best TF kernel for signals with axially distributed auto-terms \cite{BoasDSP15, boas15, Abed12}, and it facilitates independent controlling of the shape and size according to the \emph{a priori} knowledge on the signal structure. An automatic parameter optimization approach for the pre-filtering ECSK kernel is also proposed in this paper. As a result, the cleanest possible running AF is delivered to the AOK optimization process, so that the resultant adaptive kernel design significantly improves the desired TF filtering performance. (b) Unlike in \cite{liu16c2} where the structure priors for BCS-based TF reconstruction were designed based on a fixed three-category pattern, in this paper we associate the hyper-parameters with a nonlinear relationship of the TF structure. The modified structure prior is designed to more adequately model the diversified relationship with neighboring TF entries.

\subsection{Computational Complexity}

In this subsection, we analyze the computational complexity of the proposed scheme and compare it with the existing approaches for FH spectrum estimation. Three methods are compared, namely, the STFT, sparse linear regression (SLR) \cite{ange10, ange13}, and sparse Bayesian learning (SBL) \cite{zhao15} based approaches. Note that, although the terms SBL and BCS are used interchangeably for the same algorithm in the literature, we use SBL and BCS in the sequel to respectively refer to the algorithms developed in the linear and bilinear TF frameworks for convenience.

Let $\zeta$ be the length of the short-time slide window. The computational complexity of the STFT-based method is $O(N\log_2\zeta)$, which is the least among all the existing approaches. In comparison, the complexity of the SLR-based method is $O(N^2L^2)$ \cite{ange10}, where $L$ is the number of frequency bins. Similar to the STFT-based method, the SBL approach \cite{zhao15} also partitions the input signals into $P$ overlapped segments through a sliding window. The computational complexity of the SBL approach is then $O(P\zeta^3+Kg^3))$ \cite{zhao15}, where $K$ is the number of latent parameters, which is normally truncated to a value close to $P$ for a tractable Bayesian inference, and $g$ denotes the cardinality of the sampled time set in the temporal kernel basis vector, which is typically smaller than $P$. As stated in \cite{zhao15}, the computational complexities of both linear TF based methods \cite{ange10, ange13, zhao15} are actually in a very similar order. In our proposed scheme, the complexities of the pre-filtering parameter optimization, pre-filtering plus AOK processing, as well as BCS reconstruction stages are $O(GQ^2)$ \cite{Sand09}, $O(NL\log_2L)$ \cite{jone95}, and $O(N^3)$ \cite{tipp01}, respectively, where $G$ is the total number of generations, and $Q$ is the dimension of the problem, i.e., the number of the parameters to be optimized. When considering the overall computational complexity, which includes multiple terms, its order is determined by that of the fastest growing term (with the highest order of $N$). As such, the overall asymptotic computational complexity of the proposed scheme is $O(N^3)$. As such, the computational complexity of the proposed method is much higher than the STFT-based method, but is only slightly higher than the SLR and SBL approaches. This is the price we pay in order to achieve robust and accurate spectrum estimation with missing observations, as we will demonstrate in the next Section.

\section{Numerical Experiments and Analysis}
\label{s:nea}

In this section, numerical experiments are conducted to evaluate the performance of the proposed algorithm in comparison with those reported in the literature. In this section, the input SNR is defined as \cite{ange10, zhao15}
\setcounter{equation}{39}
\begin{equation}\label{eq:SNRdef}
{\rm{SNR}} \triangleq  10\,{\log _{10}}\left( {\frac{{\left\| {\bf{x}} \right\|_2^2}}{{N{\sigma ^2}}}} \right),
\end{equation}
where ${\bf{x}}$ is the signal vector, and ${\sigma ^2}$ denotes the power of additive white Gaussian noise.

\begin{figure}[!b]
\centering
\subfloat[]{\includegraphics[scale=0.63]{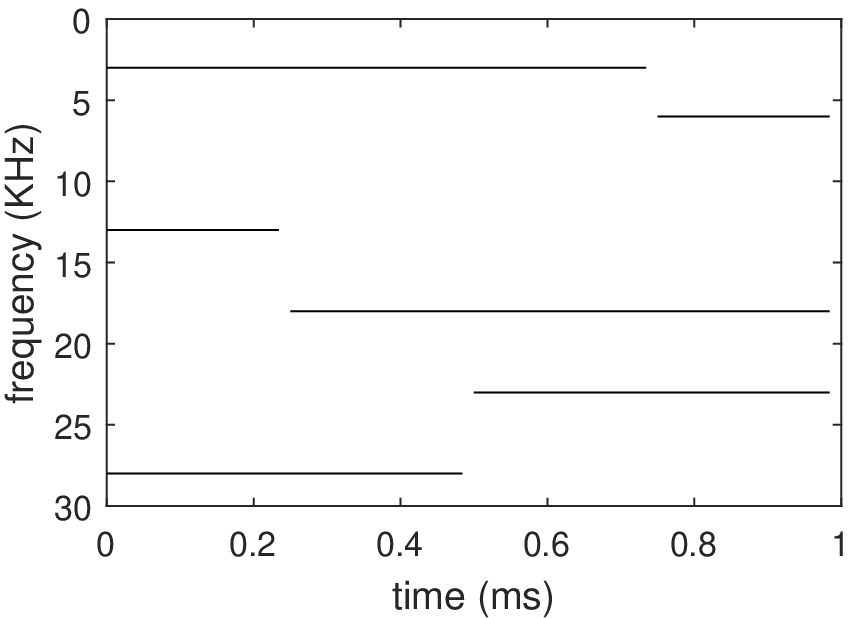}}
\hfil
\subfloat[]{\includegraphics[scale=0.63]{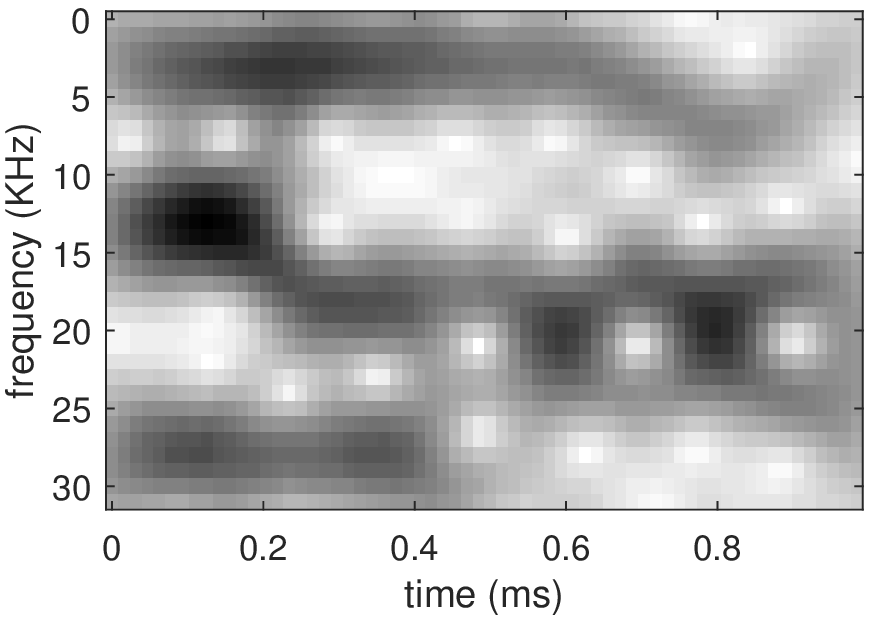}}
\caption{Simulated FH signals. (a) True TF trajectories of the FH signals; (b) Spectrogram of FH signals with missing samples.}
\label{fig:truth}
\end{figure}

In particular, two performance measures are defined for the evaluation of the hopping time and the instantaneous frequency (IF) detection performance, respectively. The ratio of correct hopping time detection is defined as \cite{zhao15}
\begin{equation}\label{eq:ptdef}
{P_t} \triangleq\frac{1}{{{M_c}}}\sum\limits_{i = 1}^{{M_c}} {{D_t}(i)}
\end{equation}
where $M_c$ is the number of Monte Carlo trails and $D_t(i)$ is the ratio of correct detections in the $i$-th Monte Carlo trial. A correct hopping time detection is declared if the estimated hopping instant is less than 3 observations away from the associated true hopping instant. The hopping time statistic is defined as ${\Delta _n} \triangleq \left\| {{x_{n + 1}} - {x_n}} \right\|_2^2$. The same definition is used in references \cite{ange10, zhao15}. The ratio of incorrect IF detection is defined as \cite{zhao15}
\begin{equation}\label{eq:efdef}
{E_f} \triangleq1-\frac{1}{{{M_c}}}\sum\limits_{i = 1}^{{M_c}} {{D_f}(i)}
\end{equation}
where $D_f(i)$ is the ratio of correct frequency detections in the $i$-th Monte Carlo trial.

\begin{figure*}[!htbp]
\centering
\subfloat[]{\includegraphics[scale=0.63]{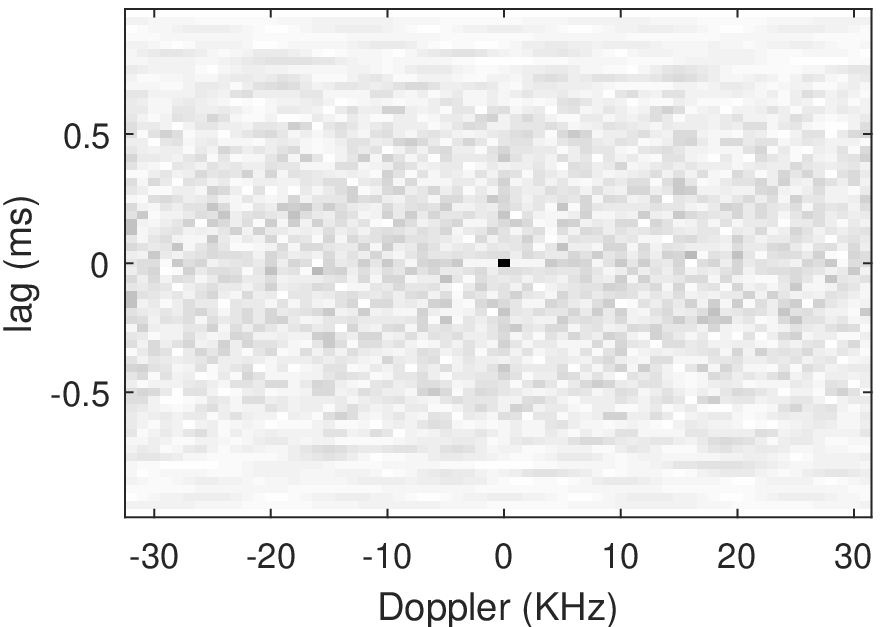}}
\hfil
\subfloat[]{\includegraphics[scale=0.63]{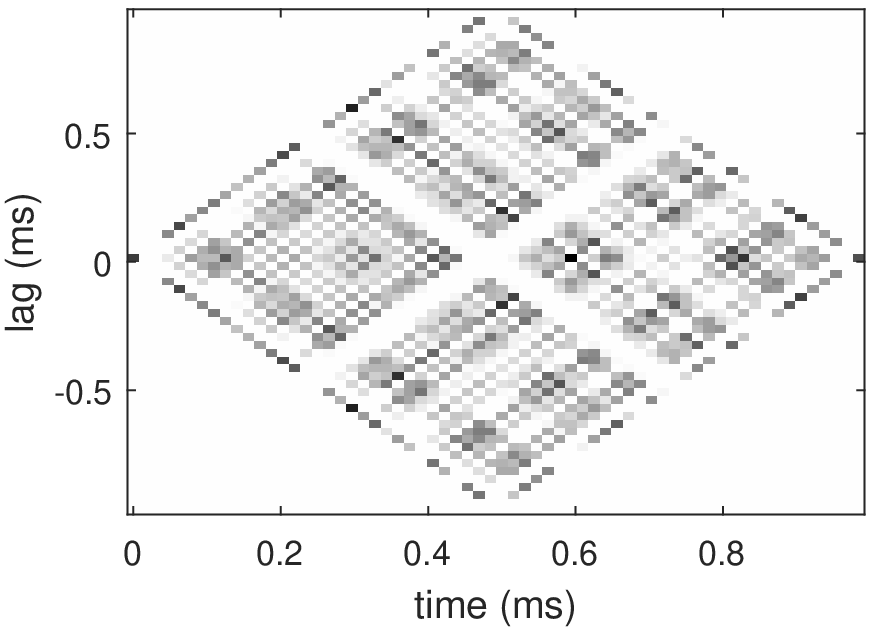}}
\hfil
\subfloat[]{\includegraphics[scale=0.63]{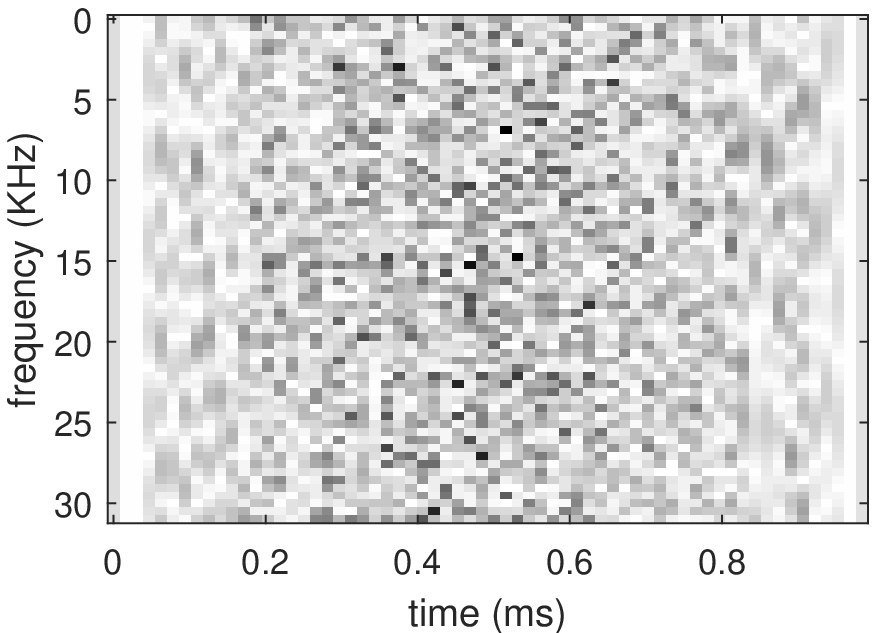}}
\hfil\\
\subfloat[]{\includegraphics[scale=0.63]{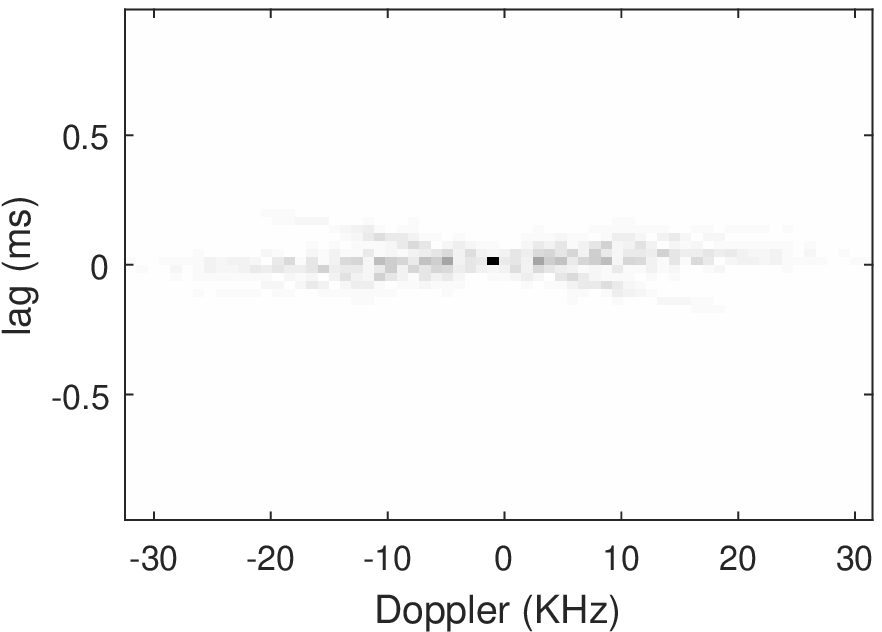}}
\hfil
\subfloat[]{\includegraphics[scale=0.63]{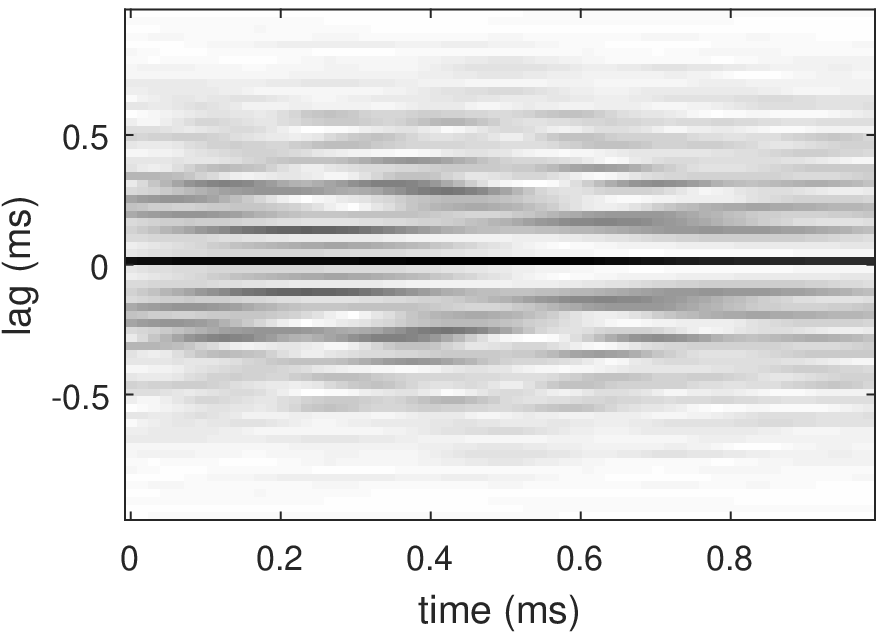}}
\hfil
\subfloat[]{\includegraphics[scale=0.63]{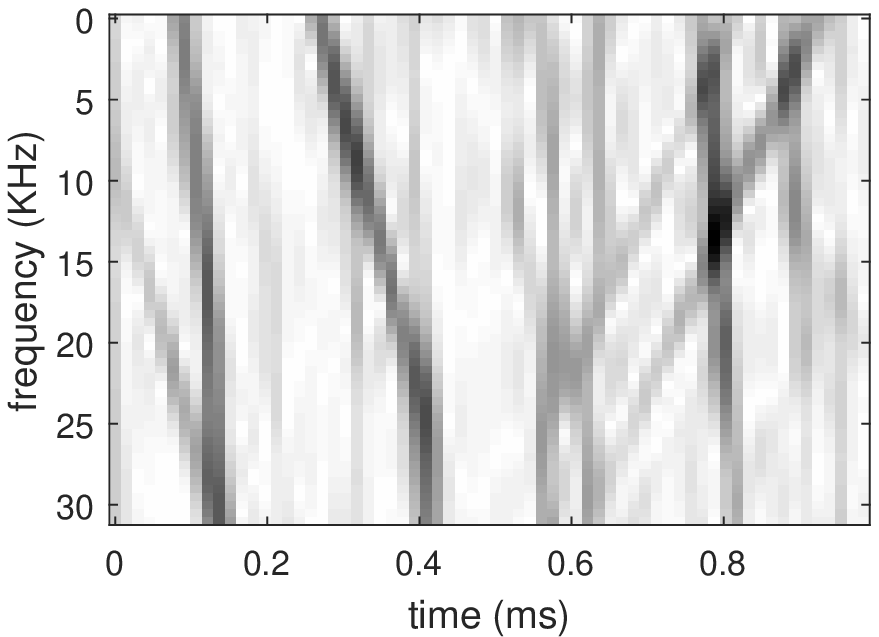}}
\hfil\\
\subfloat[]{\includegraphics[scale=0.63]{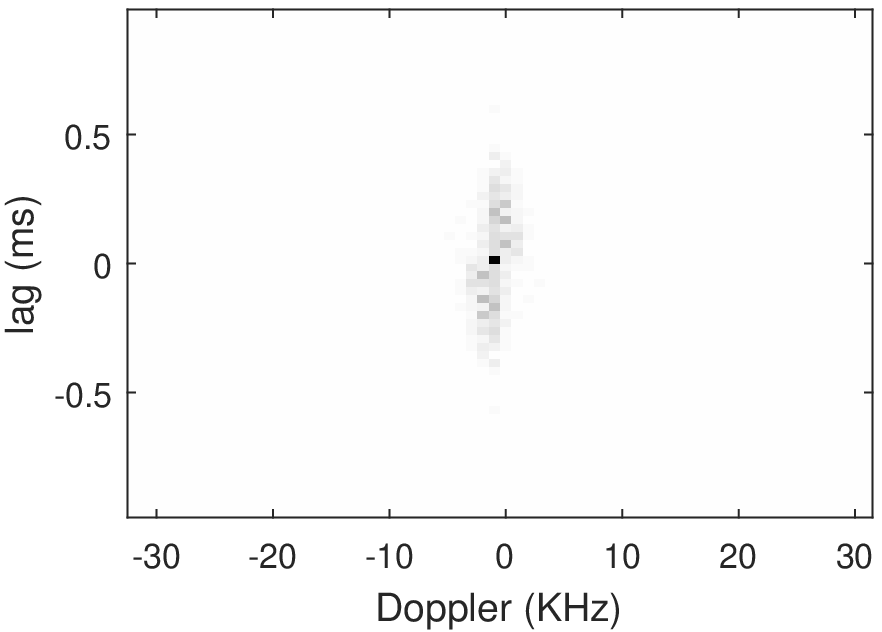}}
\hfil
\subfloat[]{\includegraphics[scale=0.63]{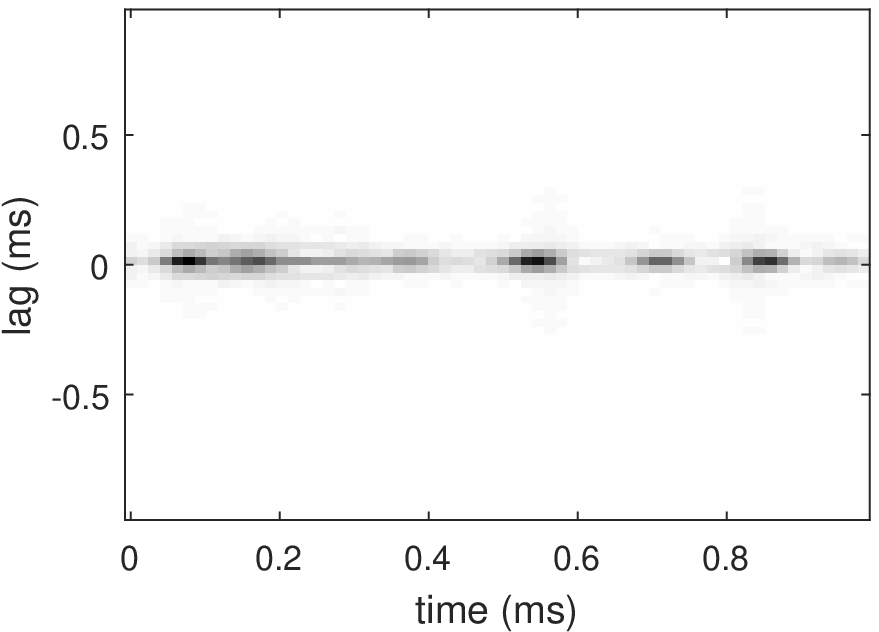}}
\hfil
\subfloat[]{\includegraphics[scale=0.63]{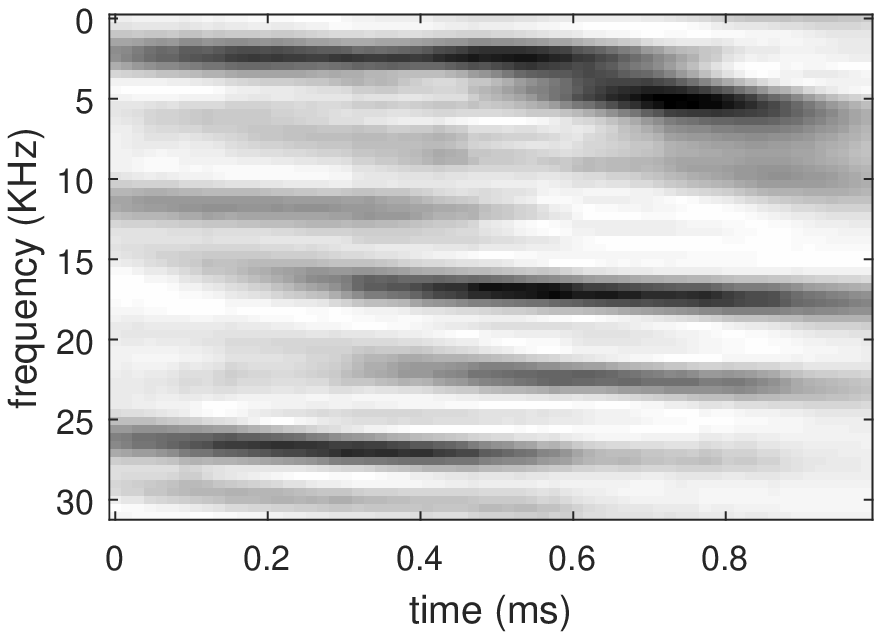}}
\caption{Joint-variable representations of FH signal with missing observations and kernel processed results. (a) AF; (b) IAF; (c) WVD; (d) AF after applying AOK; (e) IAF after applying AOK; (f) TFR after applying AOK; (g) AF after applying the proposed kernels; (h) IAF after applying the proposed kernels. (i) TFR after applying the proposed kernels.}
\label{fig:ker}
\end{figure*}

Simulation results are provided to demonstrate the effectiveness of the proposed approach. First, an illustrative example is given in Fig.\ \ref{fig:truth}(a), where the FH signals are identical to those used in reference \cite{zhao15}. The signals are generated as follows: The first FH component is active with a carrier frequency of $13$ KHz within the range of time index $[0 : 15]$ and the carrier frequency hops to $18$ KHz within the range of time index $[16 : 63]$. The second hopping component is active with a carrier frequency of $28$ KHz within the range of time index $[0 : 31]$ and the carrier frequency hops to $23$ KHz within the range of time index $[32 : 63]$. The third hopping component is active with a carrier frequency of $35$ KHz within the range of time index $[0 : 47]$ and the carrier frequency hops to $6$ KHz within the range of time index $[48 : 63]$. The sampling frequency $f_s$ is $64$ KHz. The model hyper-parameters for the priors are set as follows: $a=b=c=d=10^{-6}$, the value of $f$ is assigned as in (\ref{eq:revise_f}), and $e=1-f$. The initial conditions are set as $\alpha_i(0)=1, \pi_i(0)=0$, and $\alpha_0(0)=10^2/{\rm {var}}(\bf{c})$, where ${\rm {var}}(\cdot)$ yields the scalar variance of a vector. Fig.\ \ref{fig:truth}(a) shows the true TF trajectories of the generated FH signals. The TF analysis of such multi-component FH signals, particularly at a low input SNR, is a challenging problem. Fig.\ \ref{fig:truth}(b) shows the spectrogram with $10\%$ missing samples and input SNR of $30\;{\rm{dB}}$. It is evident that the TF signatures can be hardly recognized with linear approach even in the case where the missing-sample rate is low and the input SNR is high.

In the following, we show the superior performance achieved by the proposed method for the situation where the SNR is set to $0$ dB, and the missing-sample rate is $25 \%$. The joint-variable representations of the missing-sample FH spectrum and their kernelled versions are presented in Fig.\ \ref{fig:ker}. In Figs.\ \ref{fig:ker}(a) through \ref{fig:ker}(c), no kernel is adopted. The impact of missing samples can be clearly observed from the IAF showing in Fig.\ \ref{fig:ker}(b), and the auto-terms can hardly be identified from both AF and WVD in Figs.\ \ref{fig:ker}(a) and \ref{fig:ker}(c). Figs.\ \ref{fig:ker}(d) through \ref{fig:ker}(f) show the corresponding joint-variable representations when the AOK is applied.

\begin{figure}[!htbp]
\centering
\subfloat[]{\includegraphics[scale=0.63]{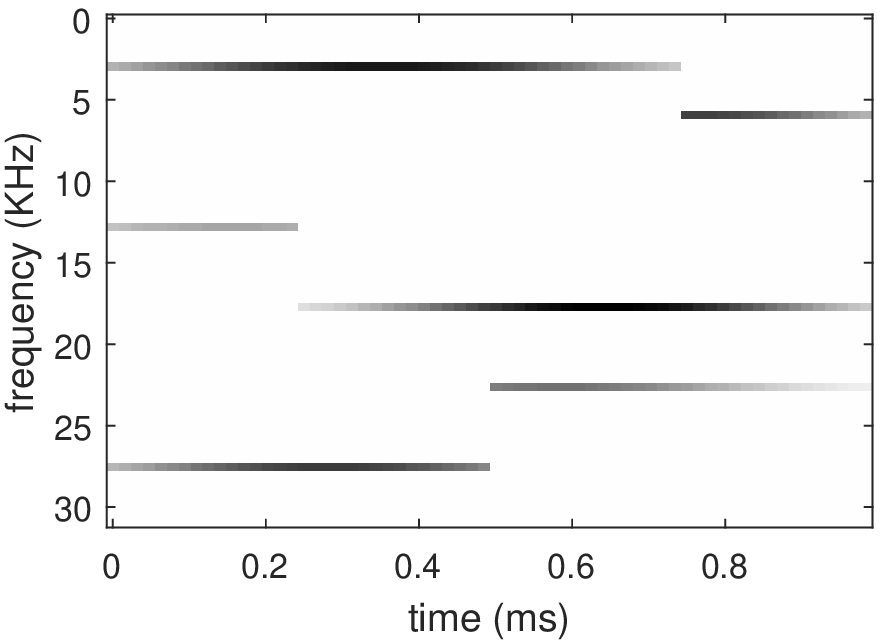}}
\hfil
\subfloat[]{\includegraphics[scale=0.63]{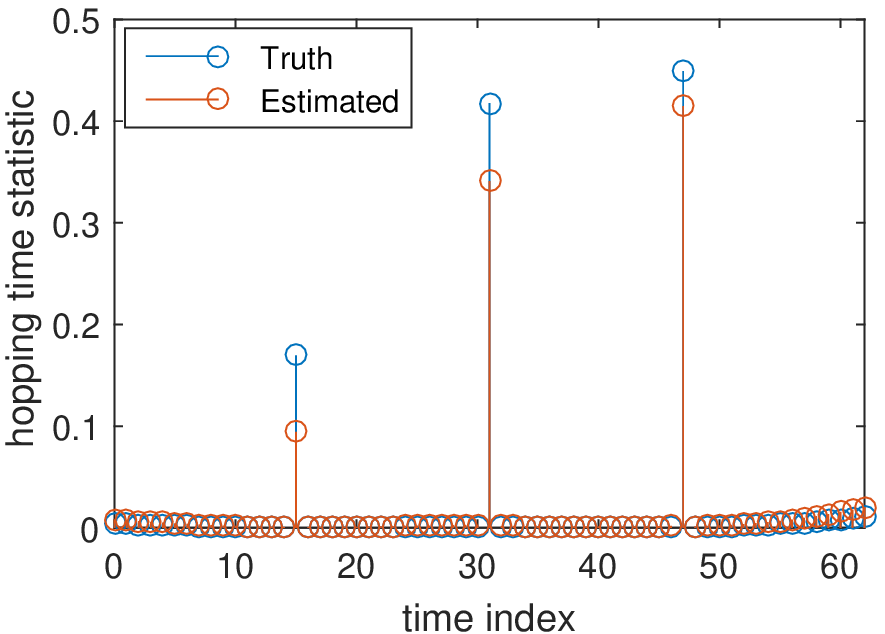}}
\caption{Estimation results: (a) Estimated TFR; (b) Hopping time statistics.}
\label{fig:esti_res}
\end{figure}

In this case, because of the low SNR and the missing samples as well as the required marginal properties, the optimization process in the AOK is severely distorted. As the result, although the AF plane is much cleaner compared with Fig.\ \ref{fig:ker}(a), a satisfactory kernelled result cannot be achieved. Rather, the estimated TFR in Fig.\ \ref{fig:ker}(f) shows strong vertical strips. In Figs.\ \ref{fig:ker}(g) to \ref{fig:ker}(i), the proposed revised ECSK plus AOK scheme is adopted. It can be observed from Fig.\ \ref{fig:ker}(g) that the auto-term energy in the AF is integrally preserved, while nearly all the undesired terms are suppressed. Nevertheless, direct estimation of the instantaneous frequencies from this plot is still difficult because of the low TF resolution. Therefore, we use the structure-aware BCS to obtain an improved FH spectrum estimation with a finer resolution. The yielding result and the comparison between true and estimated hopping time statistics are respectively depicted in Figs.\ \ref{fig:esti_res} (a) and (b), which showcase a significant improvement as compared to all the above results depicted in Fig.\ \ref{fig:ker}.

\begin{figure*}[!htbp]
\centering
\subfloat[]{\includegraphics[scale=0.72]{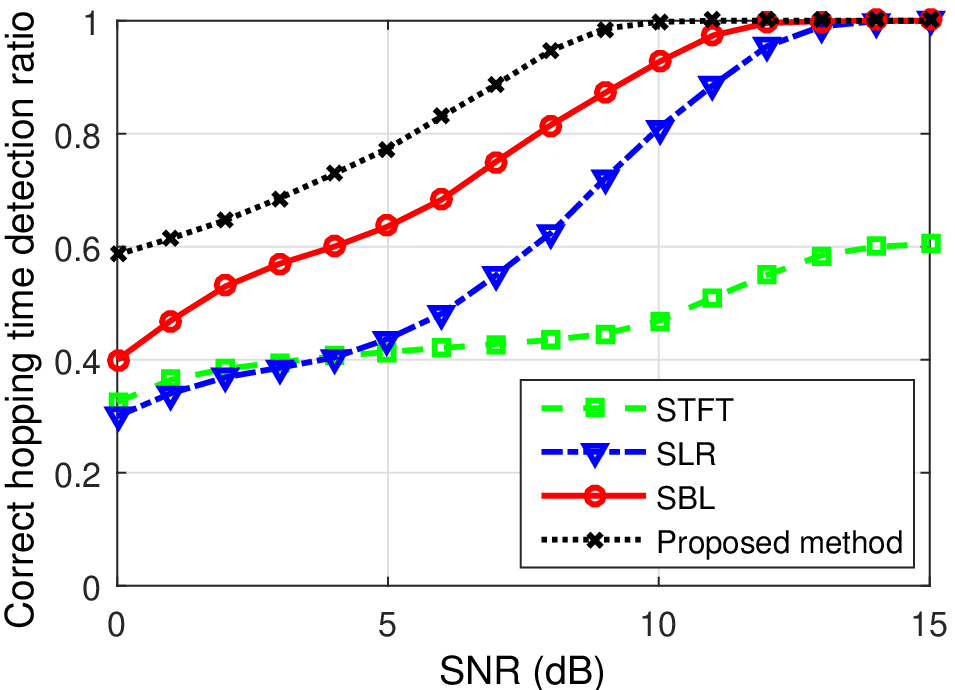}}
\hfil
\subfloat[]{\includegraphics[scale=0.72]{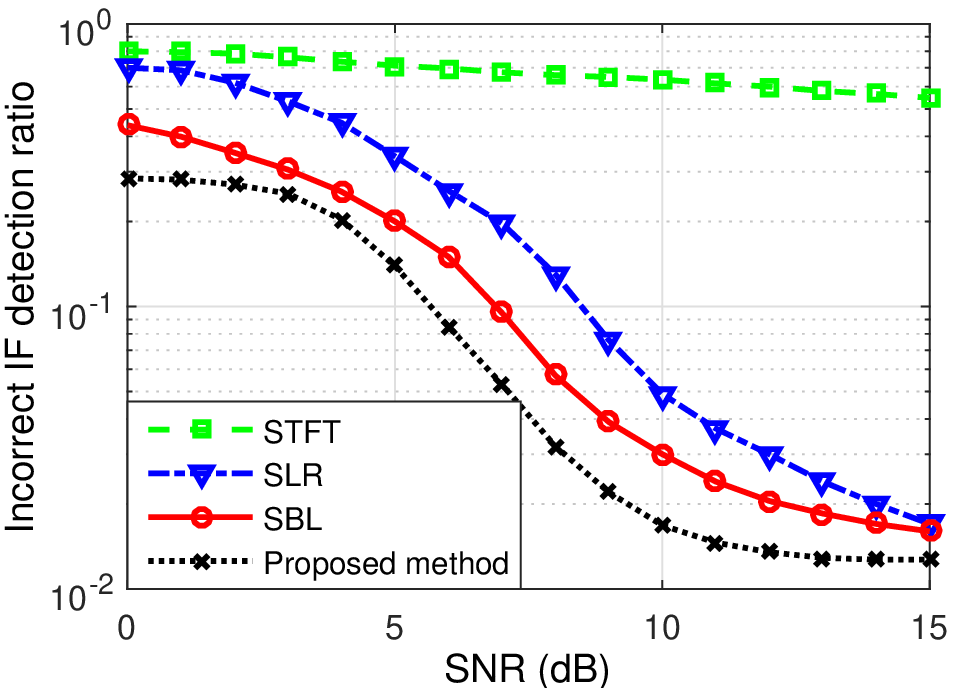}}
\caption{Statistic comparisons among different methods without missing observations. (a) Comparison of the correct hopping time detection ratio; (b) Comparison of the incorrect IF detection ratio.}
\label{fig:cp_met}
\end{figure*}

\begin{figure*}[!htbp]
\centering
\subfloat[]{\includegraphics[scale=0.72]{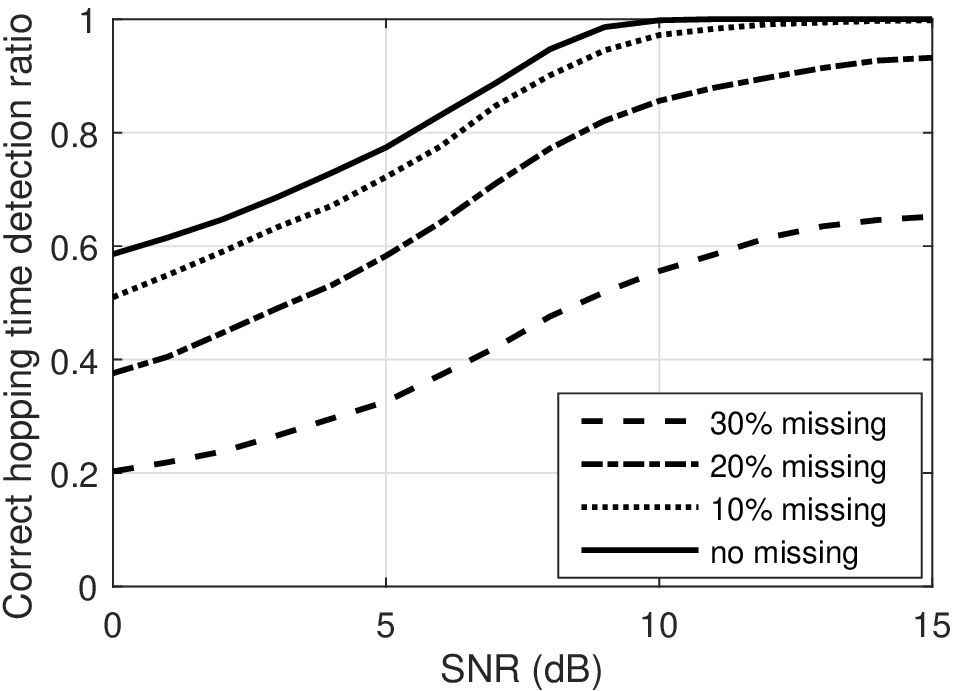}}
\hfil
\subfloat[]{\includegraphics[scale=0.72]{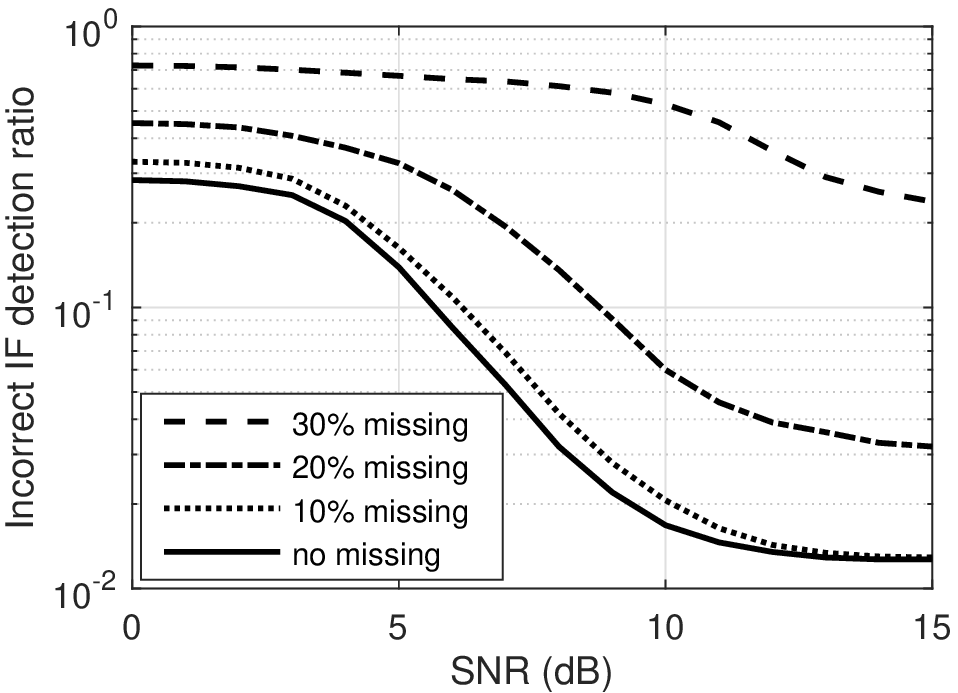}}
\caption{Statistic comparisons of the proposed method with different missing-sample rates. (a) Comparison of the correct hopping time detection ratio; (b) Comparison of the incorrect IF detection ratio.}
\label{fig:cp_mis}
\end{figure*}

\begin{figure*}[!htbp]
\centering
\subfloat[]{\includegraphics[scale=0.63]{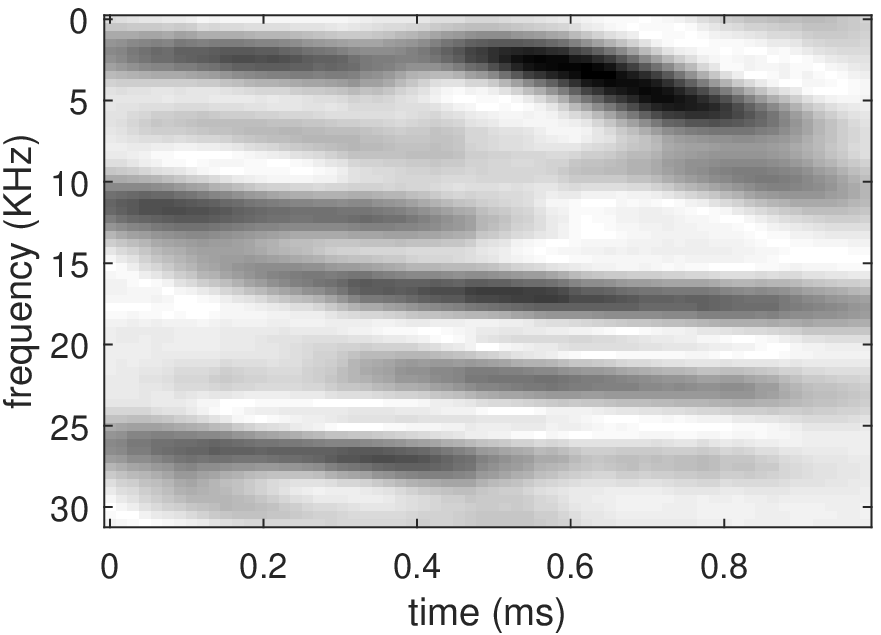}}
\hfil
\subfloat[]{\includegraphics[scale=0.63]{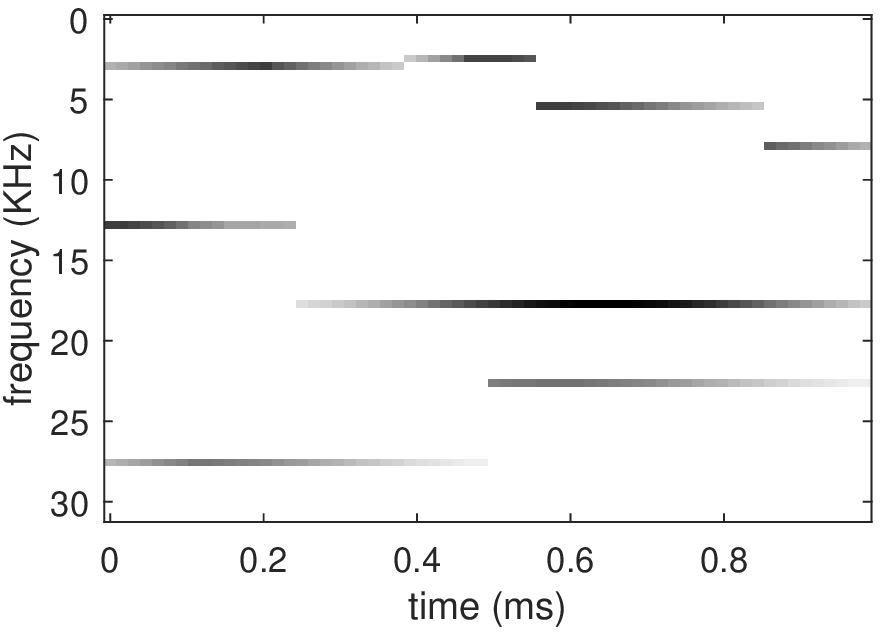}}
\hfil
\subfloat[]{\includegraphics[scale=0.63]{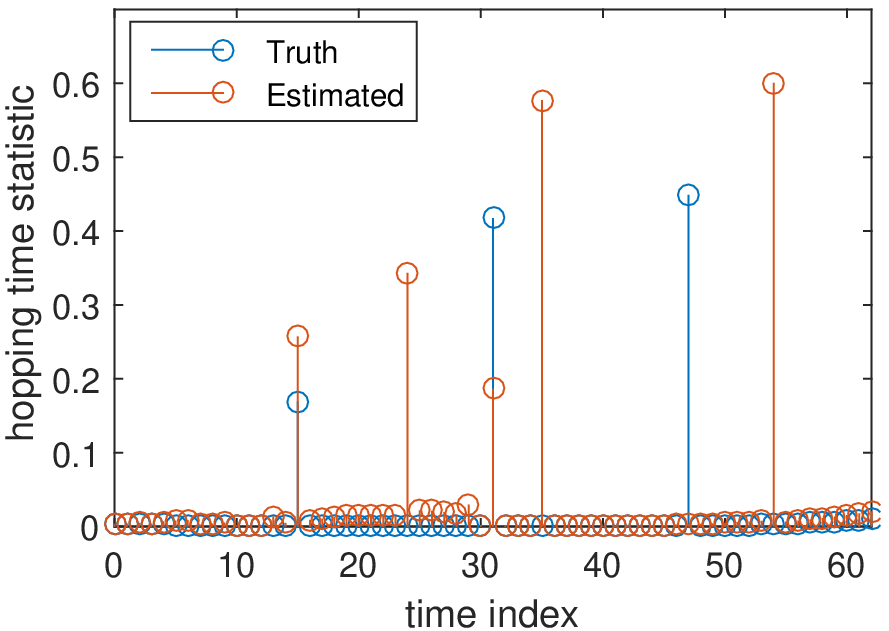}}
\caption{Impact of adopting average optimized parameters in the pre-filtering process: (a) TFR after applying the proposed kernels; (b) Estimated TFR; (c) Hopping time statistics.}
\label{fig:avrg}
\end{figure*}

To better demonstrate the effectiveness of the proposed method with statistical results, $1000$ Monte Carlo trials are conducted with the input SNRs varying from $0$ dB to $15$ dB. In Fig.\ \ref{fig:cp_met}, comparisons are drawn among different existing approaches. It can be summarized from Fig.\ \ref{fig:cp_met} that an improved performance is obtained by using the proposed method, and the advantage is more remarkable in the low SNR cases. Regarding the influence of missing samples and SNR on the algorithm performance, the statistical results are provided in Fig.\ \ref{fig:cp_mis}. Note that the detection performance of the other methods are very poor and thus are not included in Fig.\ \ref{fig:cp_mis} when we compare the performance in the presence of missing observations.  As stated above, existing linear TF analysis based approaches cannot robustly perform spectrum estimation with missing observations. Hence, in the presence of missing samples as studied in this paper, these methods yield a detection ratio which is very close to $0$ for all SNR values being investigated.

To explore the possibility to skip the pre-filtering parameter optimization process by adopting the average values after collecting sufficient estimations, we conduct numerical trials using the above simulation settings. As the simulation results shown in Fig.\ \ref{fig:avrg} indicate, this inevitably affect the pre-filtering performance and consequently slightly degrade the reconstruction accuracy.

\emph{Remarks:} Unlike the method proposed in \cite{zhao15} which considers FH signal recovery using linear TF analysis (i.e., STFT), the proposed work utilizes the bilinear TF methods. As the bilinear TF methods can use kernel designs to filter out undesired signal components, the proposed method can better utilize the known properties of FH signatures to design the kernels, thus enhancing the FH signal before applying BCS-based sparse reconstruction. This is particularly important in the presence of strong artifacts and noise. Note that the design of such kernels in the structure-aware context is a core contribution of this paper. Such kernel design is not offered in the linear STFT-based approaches. As such, the proposed work is very different to that in \cite{zhao15} and its advantages can be easily understood in concept and are clearly demonstrated through the above simulation results.

\section{Conclusion}

In this paper, a novel structure-aware FH spectrum estimation approach with the consideration of missing observations was proposed in the sparse reconstruction framework. In particular, a TF kernel was designed to effectively utilize the inherent FH signal structure. The kernelled joint-variable representation over time and lag was used to provide the TF signal representation through sparse reconstruction. In the sparsity-based spectrum estimation process, the structure of the entry under test and its neighborhood is exploited to impose a structure prior on the Bayesian inference. It was shown that this approach significantly outperforms existing approaches devised for the same problem.

\section*{Acknowledgment}
The authors would like to thank the anonymous reviewers for their valuable comments, which have helped improve the quality and clarity of this paper.


\begin{IEEEbiography}[{\includegraphics[width=1in,height=1.25in,clip,keepaspectratio]{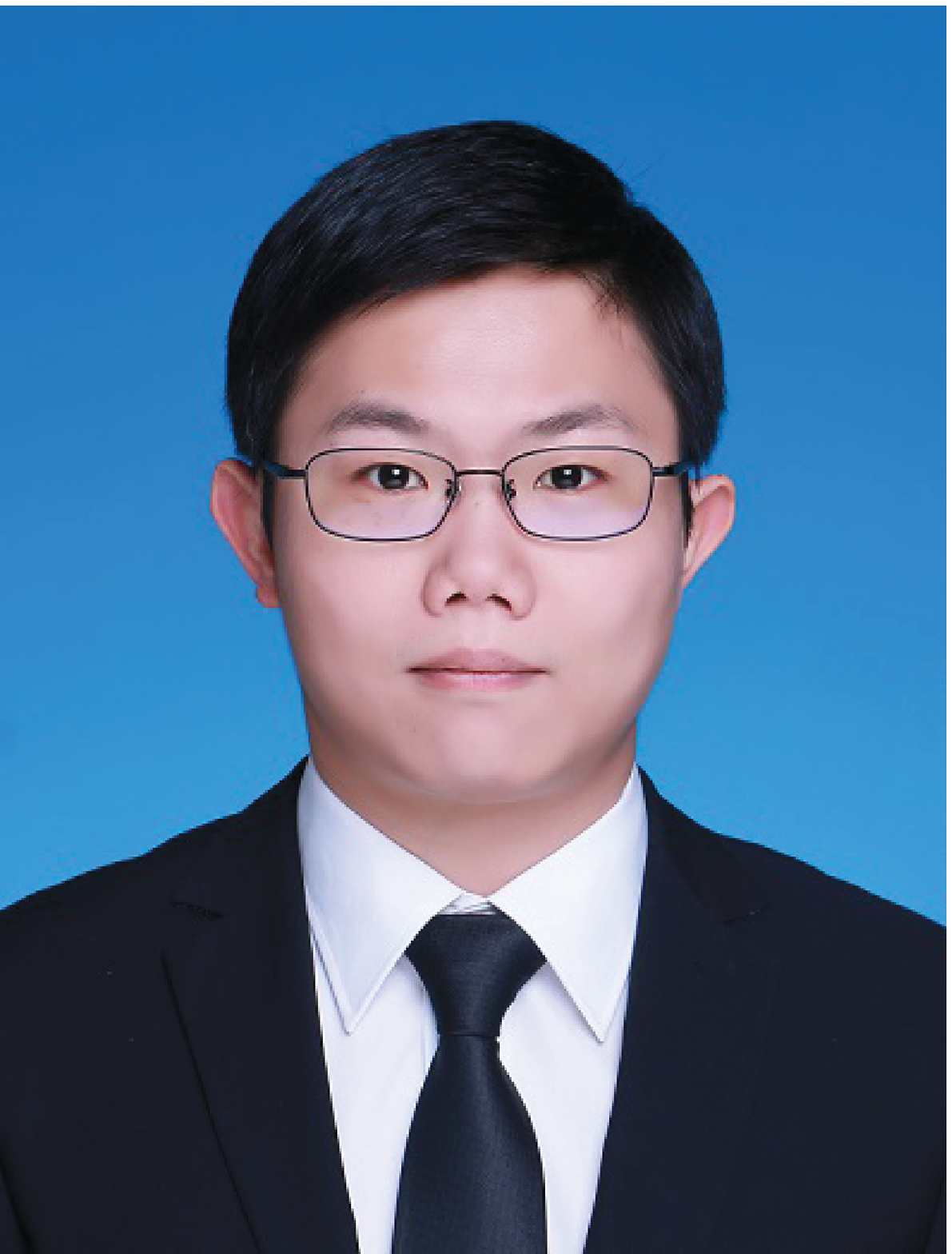}}]{Shengheng Liu}
(S'14-M'17) is currently a Postdoctoral Fellow at the Institute for Digital Communications, School of Engineering, The University of Edinburgh, UK. Prior to joining UoE, he received the B.Eng. and Ph.D. degrees in Electronics Engineering from the School of Information and Electronics, Beijing Institute of Technology, China, in 2010 and 2017 respectively. He also worked as a Visiting Research Associate from 2015 to 2016 at the Department of Electrical and Computer Engineering, Temple University, Philadelphia, PA, USA, under the support of the China Scholarship Council. His research interests include compressive sensing and time-frequency analyses for non-stationary signals, interference cancellation and coherent integration for passive bistatic radars, as well as image reconstruction in electrical impedance tomography. He is a frequent reviewer for several top-tier journals, including \textsc{IEEE Transactions on Signal Processing}, \textsc{IEEE Transactions on Audio, Speech, and Language Processing}, and \textsc{IEEE Transactions on Instrumentation and Measurement}.
\end{IEEEbiography}

\begin{IEEEbiography}[{\includegraphics[width=1in,height=1.25in,clip,keepaspectratio]{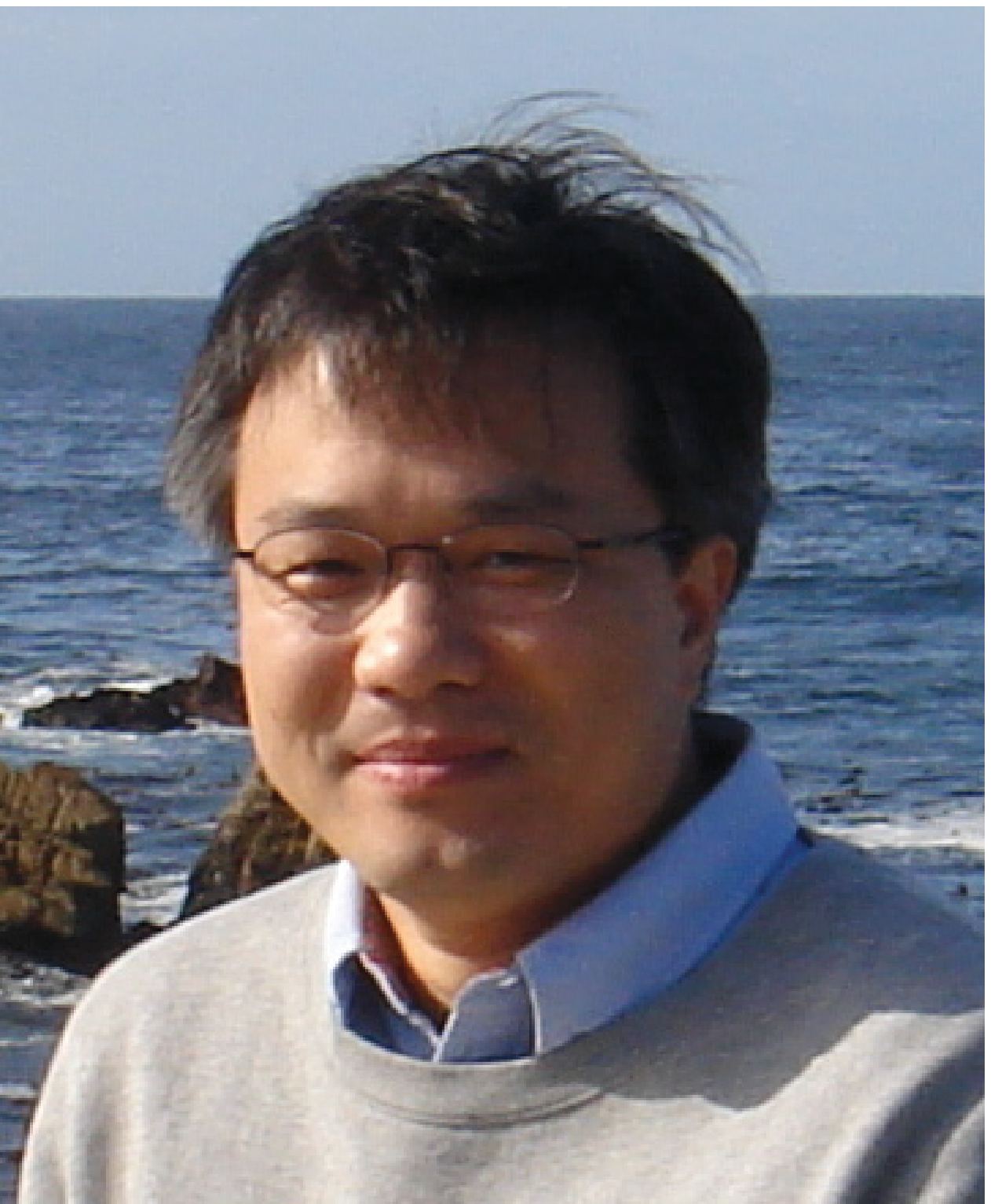}}]{Yimin D. Zhang}
(SM'01) received his Ph.D. degree from the University of Tsukuba, Tsukuba, Japan, in 1988.
	
He joined the faculty of the Department of Radio Engineering, Southeast University, Nanjing, China, in 1988. He served as a Director and Technical Manager at the Oriental Science Laboratory, Yokohama, Japan, from 1989 to 1995, and a Senior Technical Manager at the Communication Laboratory Japan, Kawasaki, Japan, from 1995 to 1997. He was a Visiting Researcher at the ATR Adaptive Communications Research Laboratories, Kyoto, Japan, from 1997 to 1998. From 1998 to 2015, he was with the Villanova University, Villanova, PA, where he was a Research Professor at the Center for Advanced Communications, and was the Director of the Wireless Communications and Positioning Laboratory and the Director of the Radio Frequency Identification (RFID) Laboratory. Since August 2015, he has been with the Department of Electrical and Computer Engineering, College of Engineering, Temple University, Philadelphia, PA, where is an Associate Professor. His general research interests lie in the areas of statistical signal and array processing for radar, communications, and satellite navigation applications, including compressive sensing, convex optimization, nonstationary signal and time-frequency analysis, MIMO systems, radar imaging, target localization and tracking, wireless and cooperative networks, and jammer suppression.

He has 12 book chapters and more than 300 journal articles and peer-reviewed conference papers. Dr. Zhang is an Associate Editor for the \textsc{IEEE Transactions on Signal Processing}, and an Editor for the \emph{Signal Processing} journal. He was an Associate Editor for the \textsc{IEEE Signal Processing Letters} during 2006--2010, and an Associate Editor for the \emph{Journal of the Franklin Institute} during 2007--2013. Dr. Zhang is a member of the Sensor Array and Multichannel (SAM) Technical Committee of the IEEE Signal Processing Society, and a Technical Co-chair of the 2018 IEEE Sensor Array and Multichannel Signal Processing Workshop.
\end{IEEEbiography}

\begin{IEEEbiography}[{\includegraphics[width=1in,height=1.25in,clip,keepaspectratio]{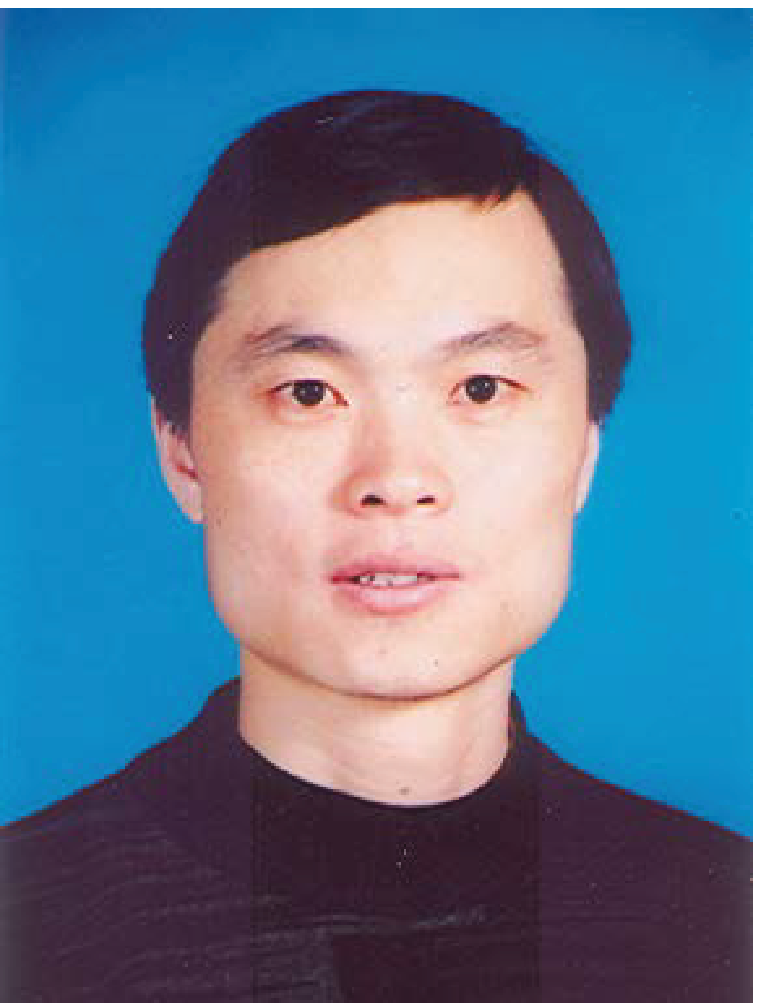}}]{Tao Shan}
(M'15) received his B.S. degree from Xidian University, Xi'an, in 1991 and Ph.D. degree from Beijing Institute of Technology in 2004. Currently, he is an Associate Professor with the School of Information and Electronics, Beijing Institute of Technology. From 2014 to 2015, he was a Senior Visiting Scholar at the Center for Advanced Communications, Villanova University, PA. He was a recipient of the first prize of science and technology progress awarded by the Ministry of Education in 2006 and 2007 respectively. His research interests include radar signal processing and time-frequency analysis for non-stationary signals.
\end{IEEEbiography}

\begin{IEEEbiography}[{\includegraphics[width=1in,height=1.25in,clip,keepaspectratio]{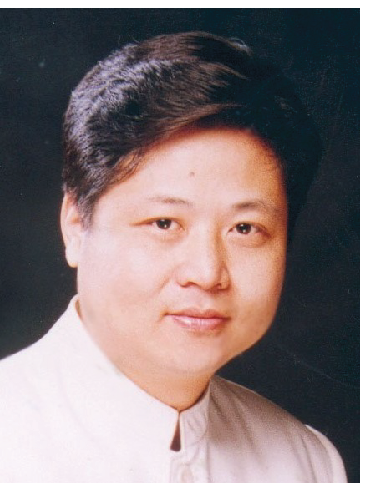}}]{Ran Tao}
(M'00-SM'04) received the B.S. degree from Electronic Engineering Institute of PLA, Hefei, in 1985 and the M.S. and Ph.D. degrees from Harbin Institute of Technology, Harbin, in 1990 and 1993, respectively. He has been a senior visiting scholar at the University of Michigan, Ann Arbor, MI, and the University of Delaware, DE, in 2001 and 2016, respectively. He is currently a Professor with the School of Information and Electronics, Beijing Institute of Technology, Beijing, China. He is a Fellow of the Institute of Engineering and Technology (IET), and a Fellow of the Chinese Institute of Electronics (CIE).
	
Dr. Tao was a recipient of National Science Foundation of China for Distinguished Young Scholars in 2006, and a Distinguished Professor of Changjiang Scholars Program in 2009. He has been a Chief Professor of the Creative Research Groups of the National Natural Science Foundation of China since 2014, and he was a Chief Professor of the Program for Changjiang Scholars and Innovative Research Team in University during 2010 to 2012. He is currently the Vice Chair of IEEE China Council. He is also the Vice Chair of the International Union of Radio Science (URSI) China Council and a Member of Wireless Communication and Signal Processing Commission of URSI. He was a recipient of the first prize of science and technology progress in 2006, 2007, respectively, and the first prize of natural science in 2013, both awarded by the Ministry of Education. His current research interests include fractional Fourier transform and its applications, theory and technology for radar and communication systems. He has 3 books and more than 100 peer-reviewed journal articles.
\end{IEEEbiography}

\end{document}